\documentclass[twocolumn]{article}
\pdfoutput=1

\usepackage{subfigure,bm,amssymb,graphicx,cite}
\usepackage{abstract}

\graphicspath{{./images/}}

\author{Ezequiel R. Soul\'{e}, ersoule@fi.mdp.edu.ar\thanks{Institute of Materials Science and Technology (INTEMA), University of Mar del Plata and National Research Council (CONICET), J.B. Justo 4302, 7600 Mar del Plata, Argentina}\and Nasser Mohieddin Abukhdeir, nasser.abukhdeir@mcgill.ca\thanks{Department of Chemical Engineering, McGill University, Montr\'{e}al, Qu\'{e}bec H3A 2B2, Canada} \and Alejandro D. Rey, alejandro.rey@mcgill.ca\footnotemark[2]
}

\title{Thermodynamics, transition dynamics, and texturing in polymer-dispersed liquid crystals with mesogens exhibiting a direct isotropic/smectic-A transition}

\begin{document}

\twocolumn[
   \maketitle               
   \begin{onecolabstract}
Experimental and modeling/simulation studies of phase equilibrium and growth morphologies of novel polymer-dispersed liquid crystal (PDLC) mixtures of PS (polystyrene) and liquid crystals that exhibit a direct isotropic/smectic-A (lamellar) mesophase transition were performed for PS/10CB (decyl-cyanobiphenyl) and PS/12CB (dodecyl-cyanobiphenyl). Partial phase diagrams were determined using polarized optical microscopy (POM) and differential scanning calorimetry (DSC) for different compositions of both materials, determining both phase separation (liquid/liquid demixing) and phase ordering (isotropic/smectic-A transition) temperatures. The Flory-Huggins theory of isotropic mixing and Maier-Saupe-McMillan theory for smectic-A liquid crystalline ordering were used to computationally determine phase diagrams for both systems, showing good agreement with the experimental results. In addition to thermodynamic observations, growth morphology relations were found depending on phase transition sequence, quench rate, and material composition. Three stages of liquid crystal-rich domain growth morphology were observed: spherical macroscale domain growth (``stage I''), highly anisotropic domain growth (``stage II''), and sub-micron spheroid domain growth (``stage III''). Nano-scale structure of spheroidal and spherocylindrical morphologies were then determined via two-dimensional simulation of a high-order Landau-de Gennes model. Morphologies observed during stage II growth are typical of direct isotropic/smectic-A phase transitions, such as highly anisotropic "batonnets" and filaments. These morphologies, which are found to be persistent in direct isotropic/smectic-A PDLCs, could provide new functionality and applications for these functional materials.
   \end{onecolabstract}
]
\saythanks

\section{Introduction}

Polymer dispersed liquid crystals (PDLCs) are functional composite materials consisting of an isotropic polymer matrix and a liquid crystalline droplet phase \cite{Drzaic1995,Drzaic2006,Jeon2007}. These functional materials are used in states where liquid crystalline ordering drives the primary function of the material as in switchable windows, displays, spatial light modulators, tunable filters and other devices \cite{Drzaic1995}. PDLC materials can also be thermally actuated to transition from opaque to transparent states, allowing for other types of applications such as thermal sensors and active elements in thermo-optical memory devices \cite{Hoppe2004}.  

The formation of PDLC materials can be driven by three phase separation mechanisms: polymerization-induced (PIPS), thermally-induced (TIPS), and solvent-induced (SIPS). Phase transition phenomena exhibited by PDLCs are complex, since a combination of phase separation (chemical de-mixing), polymerization (as in PIPS), and phase ordering (liquid crystallinity) can be involved. Thus PDLC phase transformation processes involve coupled conserved (concentration) and non-conserved (liquid crystalline) order parameters. Phase transitions involving multiple order parameters have been shown to lead to unique material architectures, such as colloidal crystals dispersed in liquid crystal matrices \cite{Das2005b}.

Under isothermal conditions and in absence of convective flow and polymerization (as in TIPS), the kinetics of the coupled order parameter process involving demixing and phase ordering is described by:
\begin{eqnarray}
\frac{\partial C}{\partial t} &=& \nabla \cdot M_C \nabla \frac{\delta F}{\delta C} \label{eqn:gen1}\\
\frac{\partial \Theta}{\partial t} &=& - M_\Theta \frac{\delta F}{\delta \Theta}  \label{eqn:gen2}
\end{eqnarray}
where $C$ is the polymer concentration, $M_i$ are mobilities, $F$ is the free energy, and $\Theta$ is the order parameter of the mesophase. Since these order parameters ($C$, $\Theta$) follow different kinetic laws, their interaction results in a rich and diverse set of anisotropic growth morphologies (below the liquid crystalline transition temperature), defect formation, and domain texturing (gradients in liquid crystal order). As demonstrated in ref.~\citenum{Das2006}, intrinsic coupling terms between order parameters in the free energy $F$ of the form:
\begin{equation} \label{eqn:couple}
F_{\Theta C} = L_C \nabla \Theta \cdot \nabla C
\end{equation}
contribute to the shaping of anisotropic domains, as well as texturing ($L_C$ is a coupling parameter). In addition, the transformation growth laws, usually given in terms of power laws $R \propto t^n$ (typically $n=\frac{1}{2}$ for conserved and $n=1$ for non-conserved under deep undercooling \cite{Balluffi2005,Sutton1995}), will exhibit non-trivial dynamics due to the coupled order parameters.

\begin{figure}
\subfigure[]{\includegraphics[width=0.8in]{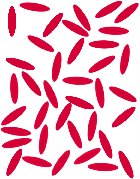}}
\subfigure[]{\includegraphics[width=0.8in]{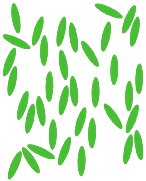}}
\subfigure[]{\includegraphics[width=0.8in]{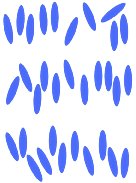}}
\caption{Schematics of the (a) isotropic (disordered), (b) nematic, and (c) smectic-A phases.}
\label{fig:lcorder}
\end{figure}

The majority of research in the field of PDLCs has focused on liquid crystals which exhibit partial orientational order \cite{Dorgan1993,Amundson1998,Borrajo1998,Lucchetti2000,Harrison2000a,Hoppe2002,Hoppe2004,Hoppe2004a,Das2006}, the nematics (see \ref{fig:lcorder}a-b). Higher-order liquid crystals also exist, such as smectic and columnar mesophases, where both orientational and translational is present (see \ref{fig:lcorder}c). The thermo-physical characterization of these higher-order liquid crystals is less complete compared to nematic liquid crystals. This situation is amplified when considering PDLCs involving smectic mesogens \cite{Benmouna1999,Benmouna2000,Graca2003} in that eqns. \ref{eqn:gen1}-\ref{eqn:gen2} have to include positional and orientational order. Additionally, when considering the formation of smectic-A mesophases, it is found that some compounds exhibit a direct isotropic/smectic-A (\ref{fig:lcorder}a $\rightarrow$ \ref{fig:lcorder}c) transition upon quenching, while others exhibit a sequential symmetry breaking cascade (\ref{fig:lcorder}a $\rightarrow$ \ref{fig:lcorder}b $\rightarrow$ \ref{fig:lcorder}c). The direct isotropic/smectic-A transition exhibited by both 10CB and 12CB and is a simultaneous symmetry breaking transition. It is of particular interest due to the complex growth morphologies \cite{Fournier1991,Todorokihara2001a} (rods, filaments, ``batonnet'' structures) and dual order parameter transformation kinetics \cite{Abukhdeir2009b} that have been observed in past studies. When considering TIPS-based PDLCs, it is expected that the additional presence of mass diffusion will expand the phenomenology already found when using nematogens \cite{Malik2006}. In particular, there will be three binary coupling free energies (as eqn. \ref{eqn:couple}, one conserved and two non-conserved), assuming isothermal conditions.
 
PIPS has been the standard approach to producing PDLCs in that the final PDLC material has a structure which is more thermodynamically stable when exposed to temperature changes and solvents \cite{Dorgan1993,Amundson1998,Borrajo1998,Lucchetti2000,Harrison2000a,Nakazawa2001,Hoppe2002,Hoppe2004,Hoppe2004a,Das2006}. In this work we use SIPS to generate the initial material, and reset the microstructure using TIPS.

The overall objective of this work is to characterize the thermodynamic and material transformation processes of novel polymer-dispersed liquid crystal (PDLC) mixtures of PS (polystyrene) and liquid crystals that exhibit a direct transition from a disordered to the lamellar smectic-A state. A comprehensive approach is taken that includes both experimental and computational techniques in order to access the multiple scales involved in PDLC phenomena. The specific objectives of this work on smectic PDLCs are divided into two areas: (1) equilibrium thermodynamics and (2) material transformation processes.

The first objective, studying equilibrium thermodynamic properties of these novel smectic PDLC materials, is accomplished by first experimentally determining a partial phase diagram. This includes determination of the isotropic/isotropic phase separation temperature ($T_{II}$) and isotropic/smectic-A phase transition temperature ($T_{AI}$), or at high concentrations of polymer the temperature at which both phase separation and ordering take place simultaneously ($T_{I/AI}$). Following this, experimental results are then compared to computationally determined phase diagrams using the Flory-Huggins theory of isotropic mixing \cite{Kurata1982} and Maier-Saupe-McMillan theory \cite{McMillan1971}, for both material systems.

The second objective involves determining phase transformation growth morphology trends based upon thermodynamic pathway, guided by the phase diagrams determined in Objective 1. A general theory identifying three distinct growth regimes (stages I, II, and III) is used based upon research in PIPS nematic PDLC systems \cite{Serbutoviez1996}. Two-dimensional simulation and scaling theories are then used, based on experimental morphology observations, to shed light on the general nano-scale structure of spheroidal and sphero-cylindrical domains.

The paper is organized as follows: the experimental procedure, models, and simulation conditions are presented, followed by the quantitative thermodynamic results (both experimental and computational). This is followed by a brief discussion of the experimentally observed phase transition behavior, referring to related past work studying lyotropic liquid crystals. Qualitative trends of morphological growth are then given with supporting POM images. Following discussion of the macro-scale experimental observations, the nano-scale simulation results are presented. Finally, conclusions are made, including guidance for future work.

\section{Experimental}

\subsection{Materials} 

Mono-disperse polystyrene (PS, Polymer Source), with molar mass $M_n=20800$ and $M_w=22200$ (provided by manufacturer), was used as received. The glass transition temperature is 99$^\circ$C, and the mass density at room temperature is 1.047 $g/cm^3$. The LCs 4-decyl-4-cyanobiphenyl (10CB, Synthon) and 4-dodecyl-4-cyanobiphenyl (12CB, Synthon), were used as received. Both LCs show a isotropic/smectic transition (no stable nematic phase) at 52.4 $^\circ$C (10CB) and 59.6 $^\circ$C (12CB). The melting temperatures are 46.5 $^\circ$C (10CB) and 48.2$^\circ$C (12CB).
	
\subsection{Sample preparation} The mixtures, with different compositions of PS and liquid crystal, were dissolved in THF and stirred manually until complete dissolution. For microscopy measurements, a small amount of the mixture was cast on a
clean glass slide and the samples were left for 24 hours to allow for complete evaporation of the solvent (verified by tracking the weight of the sample). Another glass slide was put on top of the first one, and the dry sample was sandwiched between the two glass slides. For the pure liquid crystals no solvent was used. For DSC measurements, the samples were prepared by introducing approximately 20-40 mg of the solution into aluminum DSC pans and the samples were left for 24 hours to evaporate the solvent. Then the pans were covered with an aluminum top.

\subsection{DSC Measurements} DSC measurements were performed on a Perkin Elmer Pyris 1 apparatus equipped with a Perkin Elmer 2P intracooler. A heating ramp of 5$^\circ$C/min was used in the temperature range spanning from 10 to 100$^\circ$C. Two runs were performed per sample, and the data was analyzed in the second run.

\subsection{Optical Microscope Measurements}

A Leica DMLB microscope equipped with a video camera (Leica DC100) and a hot stage (Linkam THMS 600) was used. Samples were heated above the solubilization temperature, leading to a single isotropic phase, and left for approximately 20 minutes in the homogeneous isotropic state. Afterwards, the measurements were performed as follows:
\begin{itemize}
\item For the isotropic-isotropic equilibrium temperature determination, the microscope was used without polarizers (TOM); the sample was cooled at three different cooling rates (between $0.1$ and $5^\circ$C/min, depending on the sample) until the appearance of a second phase was observed at each cooling rate. The equilibrium temperature was determined by extrapolating to cooling rate zero. 
\item For the isotropic/smectic-A transition temperature, cross-polarizers were used (POM), following the same general procedure as before heating from the liquid crystalline state until the disappearance of the smectic phase was observed. 
\item In the quenching experiments, the samples were cooled at the fastest rate possible (about 20$^\circ$C/min), using liquid nitrogen as a coolant. Two types of quenches were performed: one-step (direct quench from homogeneous to smectic-isotropic temperature region), and two step (one quench to the isotropic-isotropic region, hold for 10 minutes, and then quench to isotropic-smectic region).
\end{itemize}

\section{Modeling and Simulation} \label{sec:models}

\subsection{Thermodynamic Model} \label{sec:thermo_model}

The theoretical approach used to computationally determine the smectic PDLC phase diagrams follows that presented by Benmouna et al\cite{Benmouna2000} utilizing the Flory-Huggins theory of isotropic mixtures \cite{Kurata1982} extended to account for the presence of smectic liquid crystal phase-ordering using the Maier-Saupe-McMillan theory \cite{McMillan1971}.

\subsubsection{Flory-Huggins theory} 

The Flory-Huggins isotropic mixing free energy per mole of cell is \cite{Kurata1982}:
\begin{equation} \label{eqn:free_en_iso}
\frac{\Delta f_{mix}}{R T} = \frac{\phi}{r_{LC}} \ln{\phi} + \frac{1-\phi}{r_p}ln\left(1-\phi\right)+\chi \phi \left( 1 - \phi \right)
\end{equation}
where $R$ is the universal gas constant, $r_{LC}/r_p$ is the ratio of the volume of liquid crystal/polymer with respect to the reference volume (taken as the volume of the smaller species present, in this case a repetitive unit of the polymer), $\chi$ is the mixing interaction parameter, $\phi$ is the volumetric fraction of liquid crystal (LC), and $T$ is the temperature. The interaction parameter is considered a function of temperature only, given by $\chi = A + B/T$, and can be determined by fitting the experimentally determined $T_{I/I}$ values.

\subsubsection{Maier-Saupe-McMillan theory}

The equilibrium free energy contribution from smectic ordering according to Maier-Saupe-McMillan theory is\cite{McMillan1971}:
\begin{equation} \label{eqn:free_en_lc}
\frac{\Delta f_{ord}}{R T} = \frac{\phi}{r_{LC}} \left( \frac{1}{2} \nu \left( s^2 + \xi \sigma^2 \right) \phi -\ln{Z}\right)
\end{equation}
where $\nu=4.54 T_{NI}/T$ (see eqn. 17 of ref.~\citenum{McMillan1971}), $s/\sigma$ is the nematic/smectic order parameter, and $\xi$ is the smectic interaction parameter. The value of this parameter determines the ratio $T_{NI}/ T_{AI}$, and the order (first or second) of nematic/smectic transition\cite{McMillan1971}. $Z$ is the partition function and is given by\cite{McMillan1971}:
\begin{equation}
Z = \int_0^1 \int_0^1 exp\left[ \frac{1}{2} \nu \phi \left( 3 x^2 - 1\right) \left( s + \sigma \xi cos\left( {2 \pi z} \right) \right) \right] dz dx
\end{equation}
The total free energy density of the mixture is the sum of eqns. \ref{eqn:free_en_iso} and \ref{eqn:free_en_lc}.

\subsubsection{Phase diagram computation}

For a bi-phasic equilibrium at constant temperature, the equilibrium condition is given by equality of the chemical potentials of each component in both phases (represented by $\alpha$ and $\beta$).

\begin{equation} \label{eqn:equilibrium}
\mu_{LC}^{\alpha} = \mu_{LC}^{\beta} ; \mu_{p}^{\alpha} = \mu_{p}^{\beta} 
\end{equation}

Equilibrium exists between either two isotropic phases (I/I), one isotropic and one smectic (I/SmA), or two isotropic and one smectic (triple point). In the isotropic phases, $\sigma = s = 0$. The convergence of numerical methods when solving eq. \ref{eqn:equilibrium} usually fails when approaching the critical point. This point can be determined using the following condition:

\begin{equation} 
\frac{\partial^2 f_{mix}}{\partial \phi^2} =\frac{\partial^3 f_{mix}}{\partial \phi^3} = 0   
\end{equation}

In the smectic phase the order parameters must minimize the free energy, thus assuming $\alpha$ is the smectic phase:

\begin{equation}
\frac{\partial f_{ord}^{\alpha}}{\partial \sigma}=0 ; \frac{\partial f_{ord}^{\alpha}}{\partial s}=0 \nonumber\\
\end{equation}

The triple point is found as the intersection between the I/I and I/SmA curves

\subsection{Material Transformation Model} \label{sec:mattrans_model}

The modeling and simulation approach used to computationally determine nano-scale structure of experimentally observed smectic morphologies uses a phenomenological Landau-de Gennes model \cite{deGennes1995} first proposed for the direct isotropic/smectic-A transition by Mukherjee, Pleiner, and Brand \cite{Mukherjee2001}. This novel model has already lead to successful predictions of experimentally observed phenomena in polymer smectogens \cite{Abukhdeir2009a} and is expected to contribute new insights into the structure of smectic PDLCs.

This model is applicable to pure component smectic-A liquid crystals and is used in this work based upon a set of experimentally supported assumptions (given in Section \ref{sec:simulation}). This spatial model accounts for smectic-A ordering and elasticity and does not take into account polymer/liquid crystal interactions, as in the thermodynamic model (Section \ref{sec:thermo_model}). An extension to this material transformation model to account polymer/liquid crystal interactions has not been formulated and is beyond the scope of this paper, but as shown later, the pure smectogenic model reveals the origin of structural features driven by smectic layering.

\subsubsection{Order Parameters}

Theoretical characterization of orientational and translational order of the smectic-A mesophase (see \ref{fig:lcorder}) is accomplished using order parameters that adequately capture the physics involved. Partial orientational order of the nematic phase is characterized using a symmetric traceless quadrupolar tensor \cite{deGennes1995}:\begin{equation} \label{eqn:nem_order_param}
\bm{Q} = S \left(\bm{nn} - \frac{1}{3} \bm{I}\right) + \frac{1}{3} P \left( \bm{mm} - \bm{ll}\right)
\end{equation}
where $\mathbf{n}/\mathbf{m}/\mathbf{l}$ are the eigenvectors of $\bm{Q}$, which characterize the average molecular orientational axes, and $S/P$ are scalars which characterize the extent to which the molecules conform to the average orientational axes \cite{Rey2002,Yan2002,Rey2007}. 


The one-dimensional translational order of the smectic-A mesophase in addition to the orientational order found in nematics is characterized through the use of primary (orientational) and secondary (translational) order parameters together \cite{Toledano1987}. A complex order parameter can be used to characterize translational order \cite{deGennes1995}:
\begin{equation} \label{eq:smec_order_param}
\Psi = \psi e^{i \phi} = A+iB
\end{equation}
where $\phi$ is the phase, $\psi$ is the scalar amplitude of the density modulation, and $A/B$ is the real/imaginary component of the complex order parameter.


\subsubsection{High-Order Landau/de Gennes Model}

A dual-order parameter Landau-de Gennes model for the first order isotropic/smectic-A phase transition is used that was initially presented by Mukherjee, Pleiner, and Brand \cite{deGennes1995,Mukherjee2001} and later extended by adding nematic elastic terms \cite{Brand2001,Mukherjee2002a}:
\begin{eqnarray} \label{eq:free_energy_heterogeneous}
f - f_0 =&\frac{1}{2} a \left(\bm{Q} : \bm{Q}\right) - \frac{1}{3} b \left(\bm{Q}\cdot\bm{Q}\right) : \bm{Q} + \frac{1}{4} c \left(\bm{Q} : \bm{Q}\right)^2   \nonumber\\
& + \frac{1}{2} \alpha \left|\Psi\right|^2 + \frac{1}{4} \beta \left|\Psi\right|^4 \nonumber\\
&- \frac{1}{2} \delta \left| \Psi \right|^2 \left(\bm{Q} : \bm{Q}\right) - \frac{1}{2} e \bm{Q}:\left(\bm{\nabla} \Psi\right)\left(\bm{\nabla} \Psi^*\right) \nonumber\\
& + \frac{1}{2} l_1 \left(\bm{\nabla} \bm{Q} \right)^2 + \frac{1}{2} l_2 \left( \bm{\nabla} \cdot \bm{Q} \right)^2 \nonumber\\
& + \frac{1}{2} l_3 \bm{Q}:\left( \nabla \bm{Q} : \nabla \bm{Q} \right) \nonumber\\ 
&+ \frac{1}{2} b_1 \left|\bm{\nabla} \Psi\right|^2 + \frac{1}{4} b_2 \left|\nabla^2 \Psi\right|^2
\end{eqnarray}
\begin{equation} \label{eq:free_energy_heterogenous_coeffs}
a =  a_0 (T - T_{NI}) ; \alpha = \alpha_0 (T - T_{AI})\nonumber 
\end{equation}
where $f$ is the free energy density, $f_0$ is the free energy density of the isotropic phase, terms 1-5 are the bulk contributions to the free energy, terms 6-7 are couplings of nematic and smectic order; both the bulk order and coupling of the nematic director and smectic density-wave vector, respectively. Terms 8-10/11-12 are the nematic/smectic elastic contributions to the free energy. $T$ is temperature, $T_{NI}$/$T_{AI}$ are the hypothetical second order transition temperatures for isotropic/nematic and isotropic/smectic-A mesophase transitions (refer to ref.~\citenum{Coles1979a} for more detail), and the remaining constants are phenomenological parameters. This free energy density expression is a real-valued function of the complex order parameter $\Psi$ and its complex conjugate $\Psi^*$, which makes it convenient to reformulate using the real and imaginary parts of the complex order parameter (see eqn \ref{eq:smec_order_param}).

The previously mentioned dynamic timescales are included in this model via the Landau-Ginzburg time-dependent formulation \cite{Barbero2000}. The general form of the time-dependent formulation is as follows \cite{Barbero2000}:
\begin{eqnarray} \label{eq:landau_ginz}
\left(\begin{array}{c}
 \frac{\partial \bm{Q}}{\partial t}
\\ \frac{\partial A}{\partial t}
\\ \frac{\partial B}{\partial t} 
\end{array}\right)
&=& 
\left(\begin{array}{c c c} 
\frac{1}{\mu_n} & 0 & 0\\ 
0 & \frac{1}{\mu_S} & 0\\ 
0 & 0 & \frac{1}{\mu_S}\end{array} \right)
\left(\begin{array}{c} -\frac{\delta F}{\delta \bm{Q}}\\ 
-\frac{\delta F}{\delta A}\\ 
-\frac{\delta F}{\delta B} \end{array}\right)\\
F &=& \int_V f dV
\end{eqnarray}
where $\mu_n$/$\mu_s$ is the rotational/smectic viscosity, and $V$ the control volume. A higher order functional derivative must be used due to the second-derivative term in the free energy equation (\ref{eq:free_energy_heterogeneous}):
\begin{equation} \label{eqpdes}
\frac{\delta F}{\delta \theta} = \frac{\partial f}{\partial \theta} - \bm{\nabla} \cdot \left(\frac{\partial f}{\partial \bm{\nabla} \theta} \right) + \bm{\nabla}\bm{\nabla} : \left(\frac{\partial f}{\partial \bm{\nabla}\bm{\nabla}\theta} \right)
\end{equation}
where $\theta$ corresponds to the order parameter.

\subsubsection{Simulation Conditions} \label{sec:simcond}

\begin{figure} 
\centering
\includegraphics[width=2in]{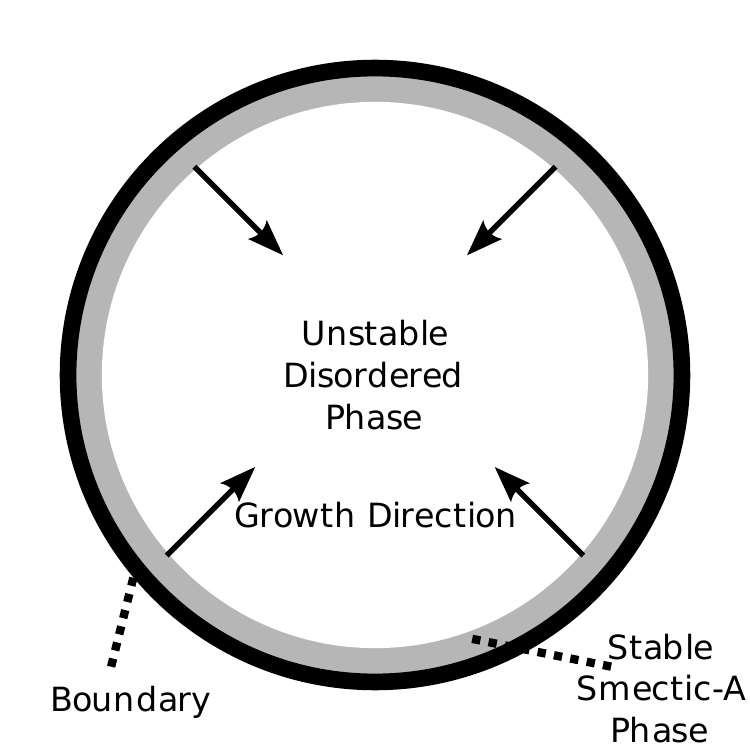}
\caption{Schematic of the two-dimensional simulation domain where the initial smectic-A nucleus growth direction into the unstable isotropic phase is shown. Dark solid lines indicate surface boundaries, grey-shaded regions indicate initially smectic-A ordered area, and white indicates initially disordered/isotropic area. The model parameters used in the simulations are as follows: $T_{NI}=322.85K$, $T_{AI}=330.5K$, $a_0= 2\times10^5\frac{J}{m^3K}$, $b=2.823\times10^7\frac{J}{m^3}$, $c=1.972\times10^7\frac{J}{m^3}$,$\alpha_0=1.903\times10^6\frac{J}{m^3 K}$, $\beta=3.956\times10^8\frac{J}{m^3}$,$\delta=9.792\times10^6\frac{J}{m^3}$, $e=1.938\times10^{-11}pN$, $b_1=1\times10^{-12}pN$,$b_2=3.334\times10^{-30}Jm$, $l_1=1\times10^{-12}pN$, and $l_2=1.033\times10^{-12}pN$. The ratio of the rotational and diffusional viscosities used was $\frac{\mu_S}{\mu_N}=25$.}
\label{fig:simschem}
\end{figure}

Two-dimensional simulation of the model was performed using the Galerkin finite element method (Comsol Multiphysics). Quadratic Lagrange basis functions were used for the Q-tensor components and quartic Hermite basis functions used for the complex order parameter components. Standard numerical techniques were utilized to ensure convergence and stability of the solution including an adaptive backward-difference formula implicit time integration method. A triangular mesh was used such that there was a density of $6$ nodes per equilibrium smectic-A layer in the circular computational domain (\ref{fig:simschem}). 

In order to simulate a boundary condition corresponding to a bulk texture, surface/boundary interaction with the liquid crystal was modeled assuming liquid crystalline ordering at the boundary fully determined by bulk conditions:
\begin{equation}
f_s = 0
\end{equation}
where $f_s$ ($J/m^2$) is the surface free energy density which is coupled to the bulk using using an elastic assumption, where the timescale of surface ordering is much greater that bulk ordering and the surface interactions are assumed always at equilibrium \cite{Barbero2000}:
\begin{equation}\label{eqn:surface_couple}
k_i \left\{ - \frac{\partial f}{\partial \frac{\partial \theta}{\partial x_i}} + \frac{\partial}{\partial x_j}\left(\frac{\partial f}{\partial \frac{\partial^2 \theta}{\partial x_i \partial x_j}} \right)\right\} = 0
\end{equation}
where $\bm{k}$ is the unit normal to the surface (outward) and $\theta$ is the order parameter. Due to the use of the full complex order parameter in the high-order model eqn \ref{eq:free_energy_heterogeneous}, boundary conditions in $\psi$ cannot be directly used. Instead, boundary conditions must be formulated in terms of $A/B$, the real and imaginary parts of the full complex order parameter. The resulting boundary conditions used in simulation include eqn \ref{eqn:surface_couple} where $\theta=A$, $k_i \frac{\partial A}{\partial x_i} = 0$, $B=0$, and $\nabla^2 B = 0$. These boundary conditions correspond to the assumption that the smectic-A phase $\phi$ at the boundary is fixed/constant. Referring back to the complex order parameter expression, this corresponds to specifying the smectic phase $\phi=(2n+1) \pi$ which fixes $A=\psi$ and $B=0$ at that point.

In this work full time-dependent simulation of the confined smectic domain and initial conditions shown in \ref{fig:simschem} is used to obtain stationary solutions to the nano-scale liquid crystalline structure. The pseudo-steady-state nano-scale structures are the main focus of these simulations and thus the dynamic growth and formation mechanisms are not discussed. This approach to stationary structure determination has a key advantage over relaxation techniques \cite{Kralj1996} in that all stationary solutions are obtained via a physically consistent pathway as predicted by the model. The main assumptions in this approach are the initial conditions (liquid crystal nucleation at the confined surfaces) and that thermal effects and flow do not play a significant role.

\section{Results and Discussion}

\begin{figure*}[htp]
\centering
\subfigure[]{\includegraphics[width=0.45\linewidth]{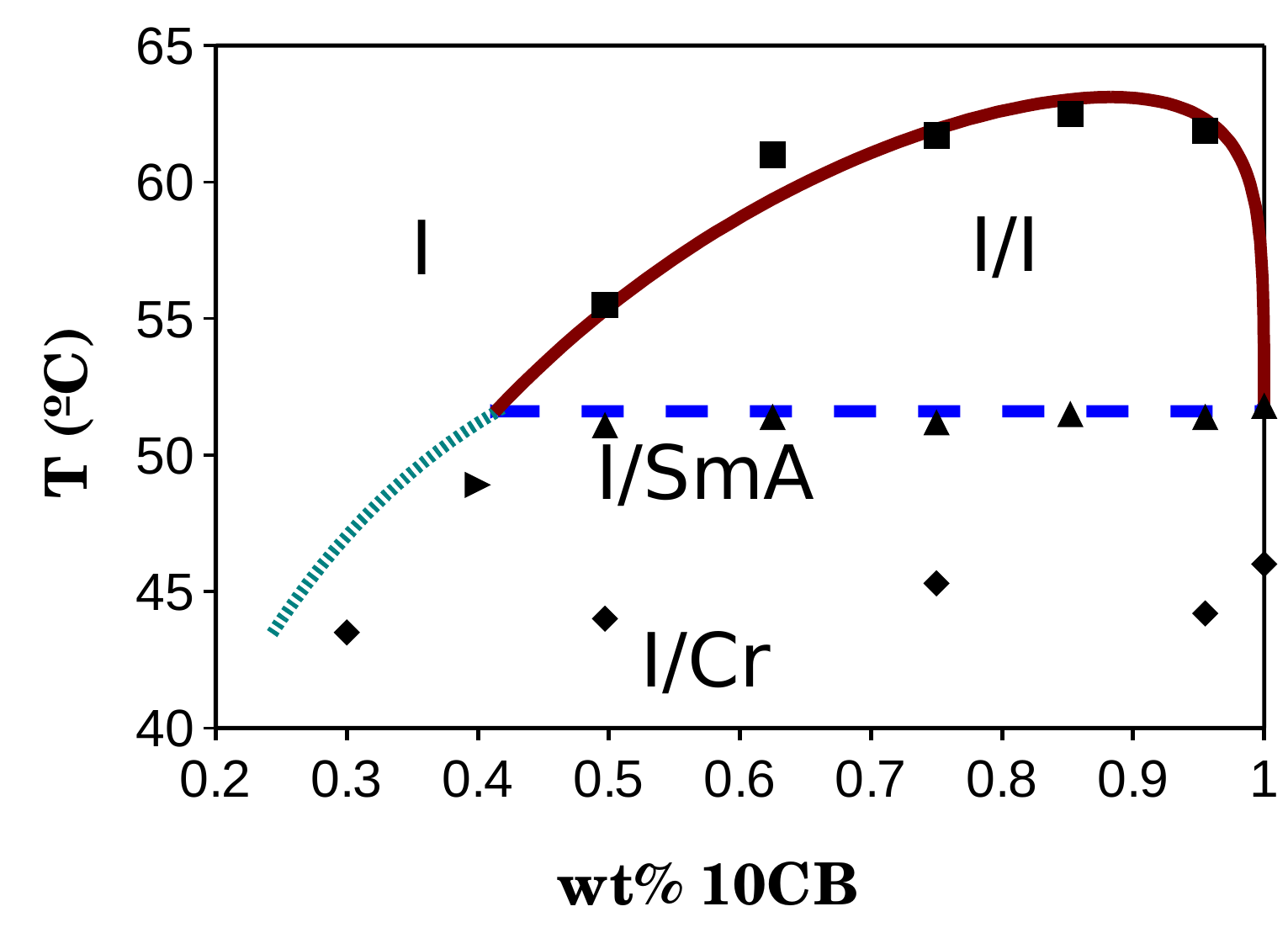}}\subfigure[]{\includegraphics[width=0.45\linewidth]{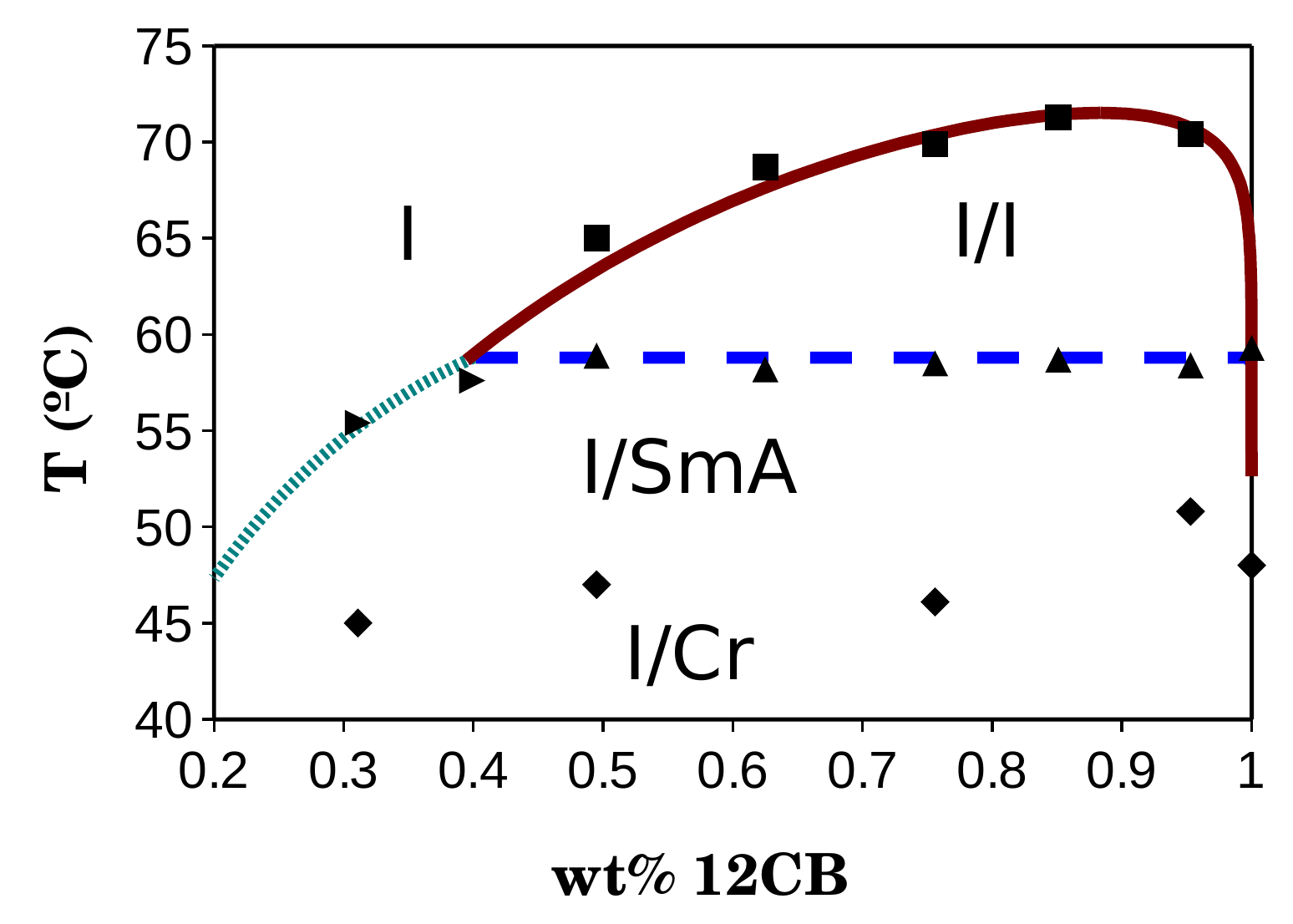}}
\caption{Experimental (points) and theoretical (lines) phase diagrams of (a) PS-10CB and (b) PS-12CB. The different regions identified in the diagrams are: homogeneous isotropic phase (I), coexistence of two isotropic phases (I/I), coexistence of a smectic and an isotropic phase (I/SmA), and coexistence of a crystalline and an isotropic phase (I/Cr). The transition from an homogeneous isotropic state to the I/I region ($\blacksquare$) was measured by TOM; the transition from I/I to I/SmA region ($\blacktriangle$) and from an homogeneous isotropic state to the I/SmA region ($\blacktriangleright$) were measured by POM and DSC, and the melting temperature ($\blacklozenge$) was measured by DSC. \label{fig:exp_phase_diag}}
\end{figure*}

\subsection{Phase diagrams}

The experimental (symbols) and theoretical (lines) phase diagrams for PS-10CB and PS-12CB, measured by POM, TOM and DSC are shown in \ref{fig:exp_phase_diag}a,b. The diagram for both mesogens shows a homogeneous isotropic phase (I), coexistence of two isotropic phases (I/I), coexistence of a smectic and an isotropic phase (I/SmA), and coexistence of a crystalline and an isotropic phase (I/Cr). Isotropic/isotropic (I/I) phase separation was determined using TOM, the isotropic/smectic-A (I/SmA) transition by POM and DSC, and crystallization by DSC only. Adequate agreement was found between the measurements by POM and DSC for the isotropic/smectic-A transition, but the DSC results showed a large dispersion, so only POM results are shown (in \ref{fig:exp_phase_diag}). For liquid crystal concentrations of 50\% in weight and greater, isotropic/isotropic phase equilibrium was observed. In this range of concentrations, the isotropic/smectic-A transition temperature and the crystallization temperature of the liquid crystal-rich phase are comparable to the pure liquid crystal values, indicating that the phase separated LC-rich domains are essentially pure liquid crystal. The increased molecular weight of the liquid crystal molecule (12CB versus 10CB) shifts the phase diagram to higher temperatures where both $T_{AI}$ and $T_{II}$ increase. 

\ref{fig:exp_phase_diag} also shows the computationally determined phase equilibrium results (full and dashed lines) using the model and experimentally determined phase diagram data. The value of $T_{N/I}$ for 10CB and 12CB (note that it is a theoretical second order isotropic/nematic temperature \cite{Coles1979a}) was obtained from literature values \cite{Coles1979a}. Smectic interaction parameters $\xi$ for both 10CB and 12CB were determined to agree with the pure component isotropic/smectic-A transition temperature (\ref{fig:exp_phase_diag}). The mixing interaction parameters $\chi = A + B/T$ were determined via least squares fitting of the experimentally determined $T_{I/I}$ and are shown in \ref{tbl:params}. The errors in the interaction parameters were obtained according to error propagation theory \cite{Draper1981,Taylor1982}. The computed phase diagram results show good agreement with the experimental data.

Experimental and calculated phase diagrams are similar to those reported by Benmouna et al. \cite{Benmouna1999,Benmouna2000} for PS/8CB (octyl-cyanobiphenyl), but with two important differences:
\begin{enumerate}
\item From the experimental point of view, their system showed the sequence isotropic/nematic/smectic, while the present system shows a direct isotropic/smectic transition. 
\item From the modeling point of view, they have three unknown, adjustable parameters  $A$, $B$ and $r_p$, while in this work an a priori value for $r_p$ is used, calculated from its definition, consequently we have only two adjustable parameters.
\end{enumerate}
The fitting method presently used also different. Benmouna et al \cite{Benmouna2000}  use the experimental critical point data to fit two of the parameters where in this work the two parameters are fit to all data points simultaneously. An interesting finding, as can be seen in \ref{tbl:params}, is that the interaction parameter can be considered independent of the molecular weight of the LC (within statistical uncertainty), which is typically not the case for low molar mass species.

\begin{table}
\caption{Material properties and model parameters used in phase diagram computation.}
\label{tbl:params}
\begin{tabular}{|c|c|c|}
\hline
\textit{parameter}&\textbf{10CB}&\textbf{12CB}\\
\hline
$M_{ref}$	& $104 \frac{g}{mol}$ 	& $104 \frac{g}{mol}$\\
\hline
$\rho_{ref}$	& $1.047 \frac{g}{cm^3}$	& $1.047 \frac{g}{cm^3}$\\
\hline
$V_{ref}$ & $99.33 \frac{cm^3}{mol}$ & $99.33 \frac{cm^3}{mol}$ \\
\hline
$M_{PS}$ &$20800 \frac{g}{mol}$&$20800 \frac{g}{mol}$\\
\hline
$\rho_{PS}$ & $1.047 \frac{g}{cm^3}$& $1.047 \frac{g}{cm^3}$\\
\hline
$V_{PS}$ & $19860 \frac{cm^3}{mol}$  & $19860 \frac{cm^3}{mol}$\\
\hline
$M_{LC}$ & $319 \frac{g}{mol}$ & $347 \frac{g}{mol}$ \\
\hline
$\rho_{LC}$ & $0.96 \frac{g}{cm^3}$ & $0.96 \frac{g}{cm^3}$ \\
\hline
$V_{LC}$ & $332 \frac{cm^3}{mol}$ & $361 \frac{cm^3}{mol}$\\
\hline
$r_p$ & $200$ & $200$ \\
\hline
$r_{LC}$ & $3.34$ & $3.63$ \\	
\hline
$T_{NI}$ &$319.6 K$& $323.6 K$ \\
\hline
$\xi$ & $1.035$ & $1.047$\\
\hline
$A$& $-1.74\pm0.19$& $-1.65\pm0.17$\\
\hline
$B$& $648\pm65$&$630\pm58$\\
\hline
\end{tabular}
\end{table}

\begin{figure}[htp]
\subfigure[]{\includegraphics[width=3in]{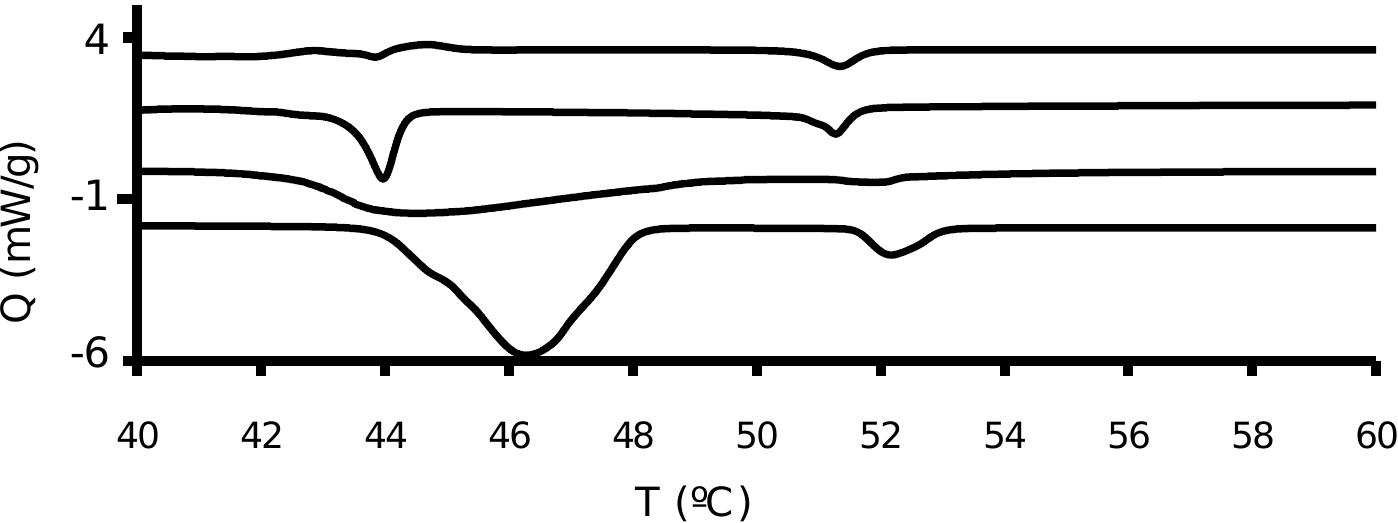}}\\
\subfigure[]{\includegraphics[width=3in]{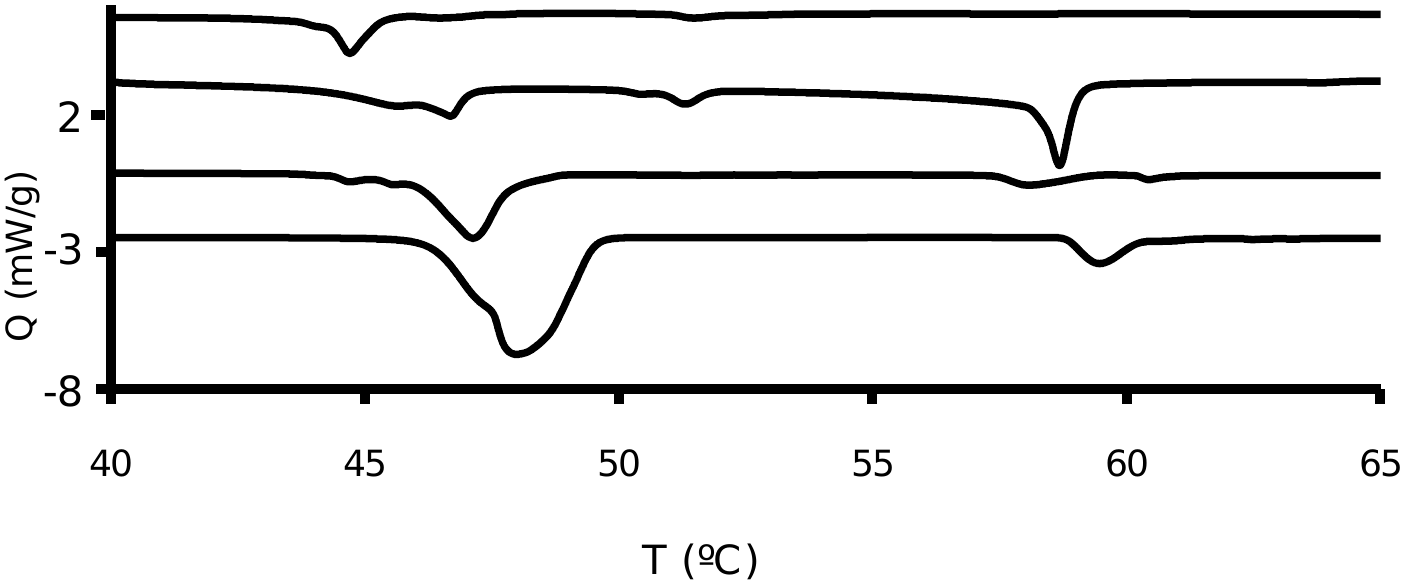}}\\
\caption{Plots of DSC data for (a) PS/10CB mixtures (from bottom to top $100.0\%$, $75.0\%$, $49.7\%$, $30.0\%$) and (b) PS/12CB mixtures (from bottom to top $100\%$, $75.6\%$, $49.5\%$, $31.1\%$). \label{fig:dsc}}
\end{figure}

\begin{figure}[htp]
\centering
\subfigure[]{\includegraphics[width=1.5in]{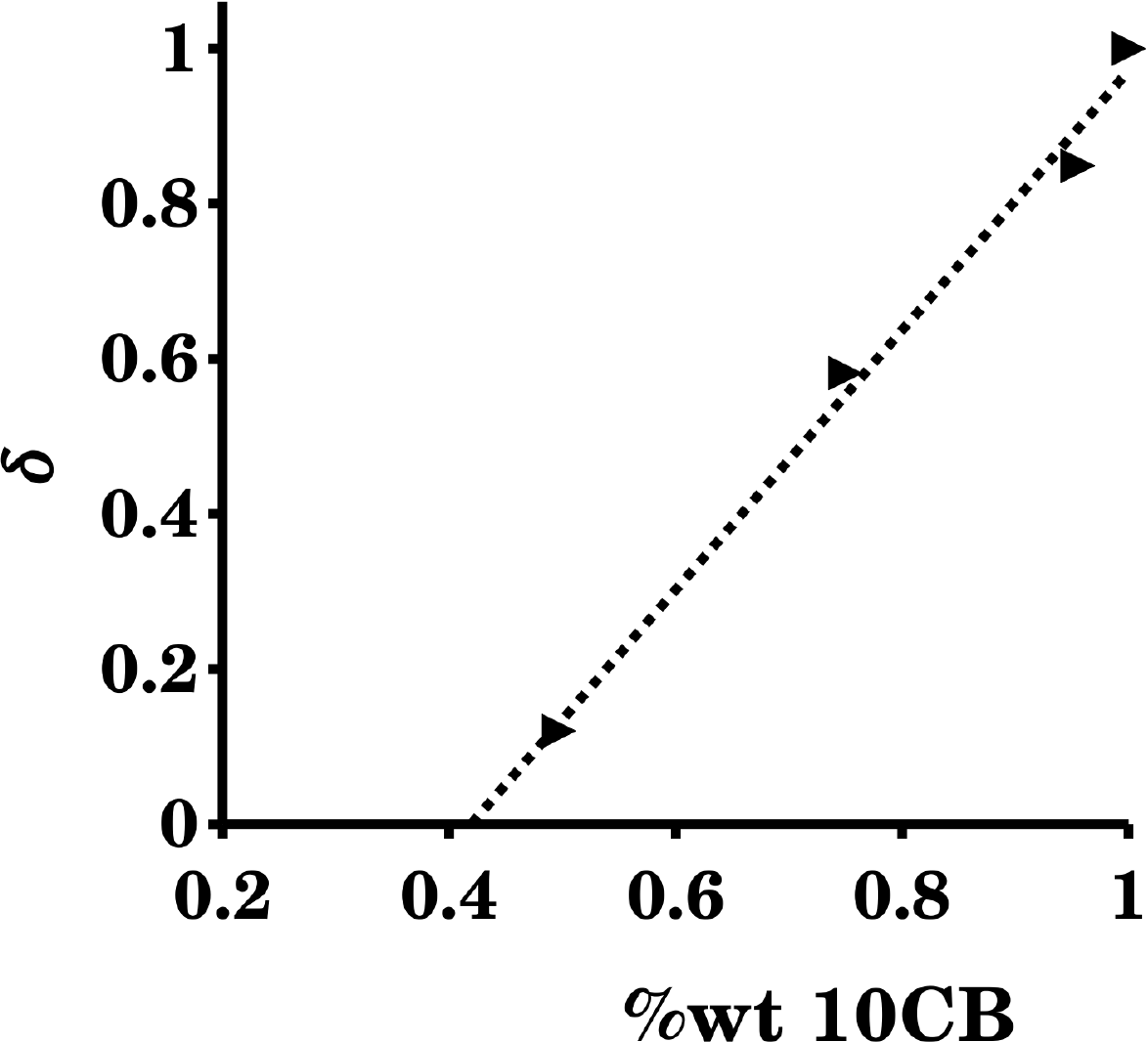}}\subfigure[]{\includegraphics[width=1.5in]{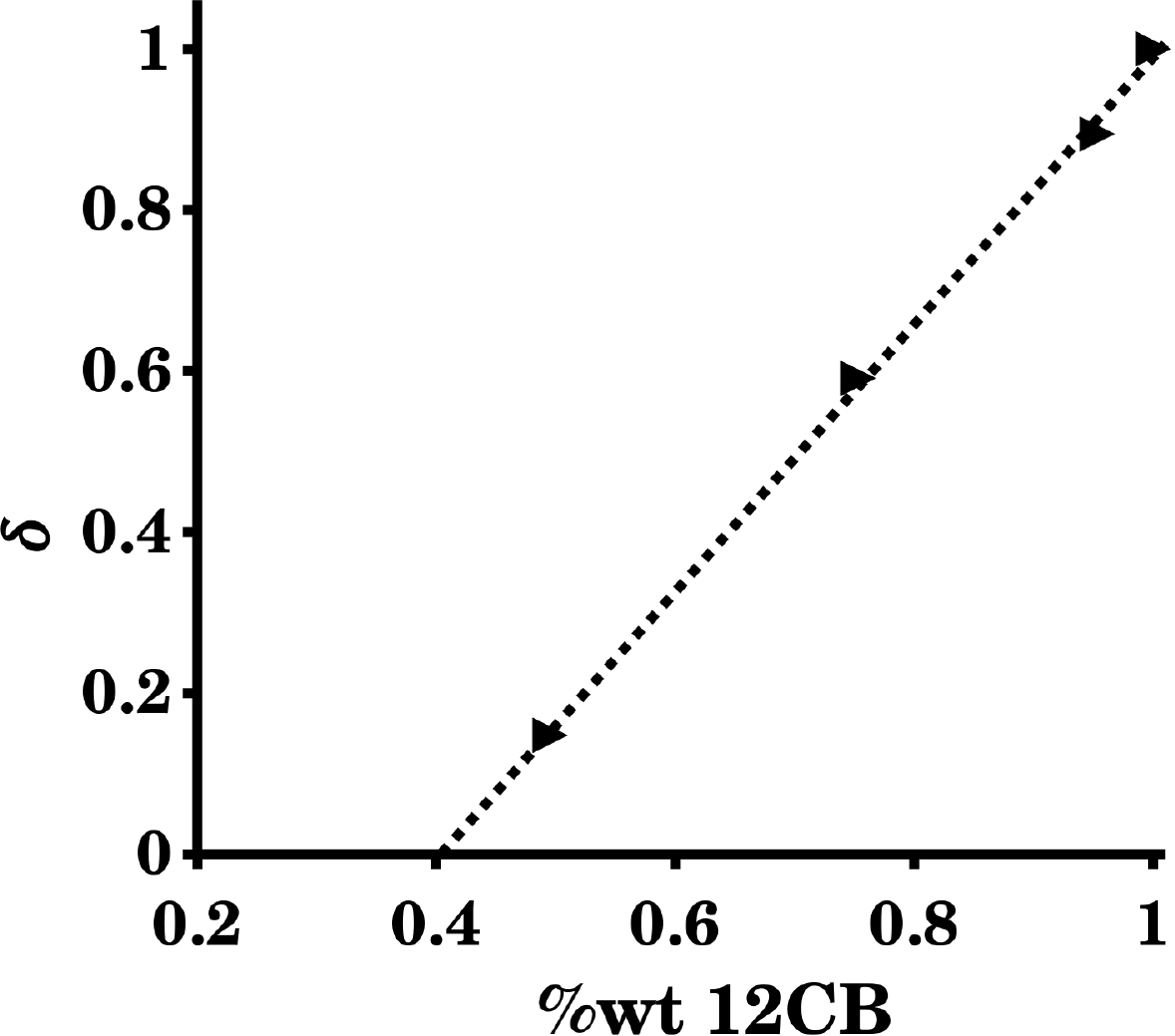}}\\
\subfigure[]{\includegraphics[width=1.5in]{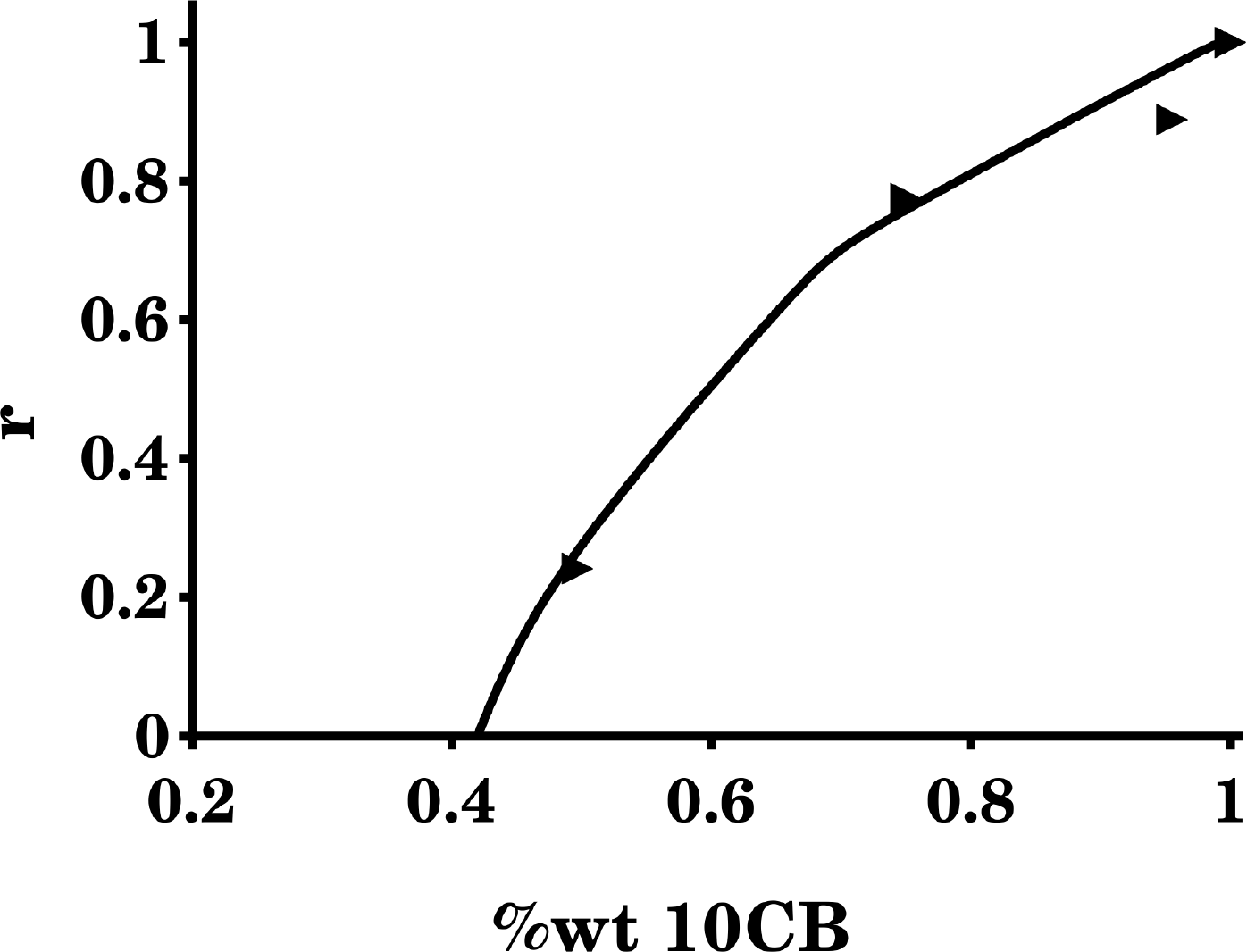}}\subfigure[]{\includegraphics[width=1.5in]{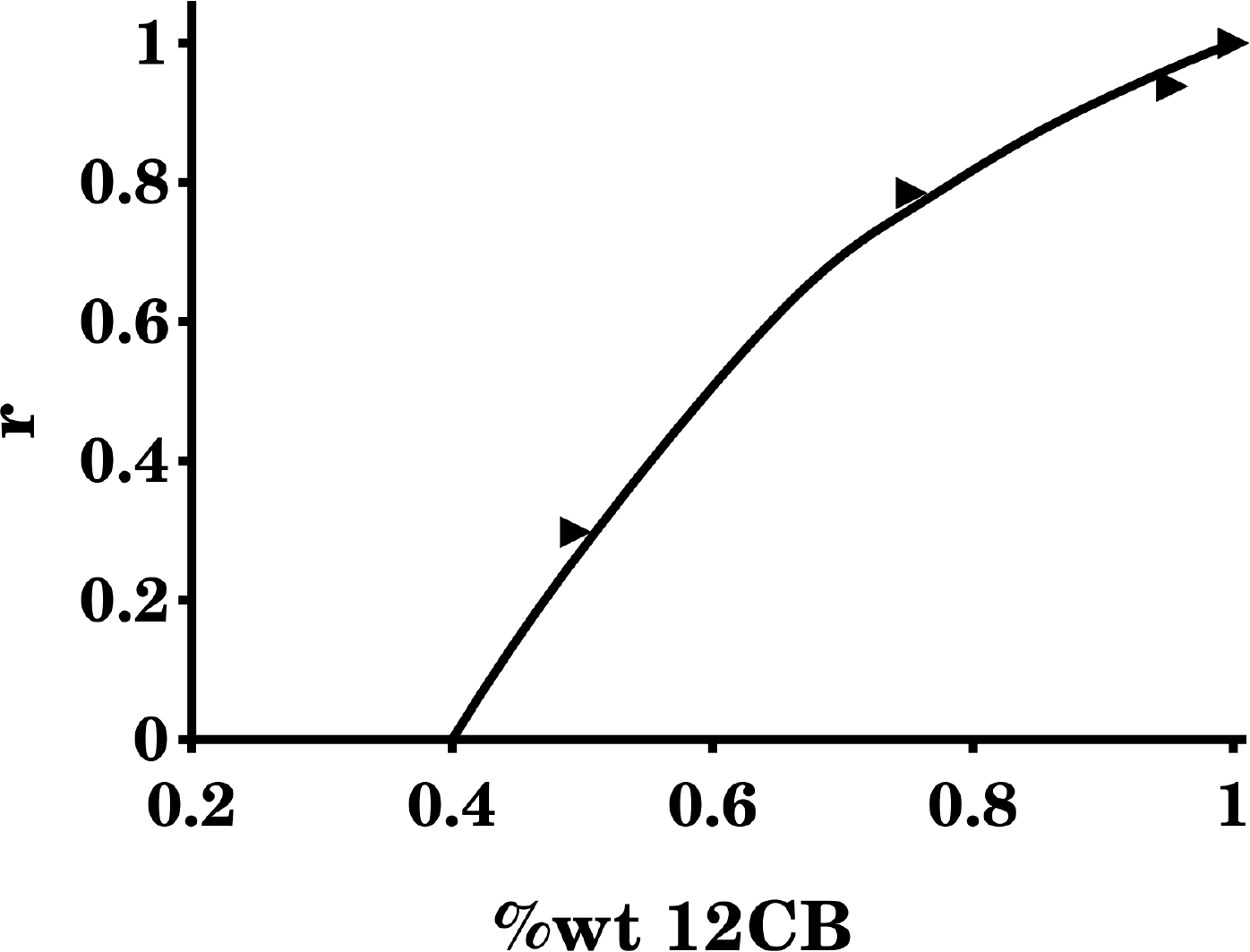}}
\caption{Plots of the enthalpy ratios versus weight percent liquid crystal for (a) PS/10CB, (b) PS/12CB, and plots of mass fraction of LC in phase-ordered (smectic-A) domains versus total mass fraction of LC for (c) PS/10CB, (d) PS/12CB. Curve fits in (a)-(b) correspond to eqn \ref{eqn:delta2} and for (c)-(d) correspond to eqn. \ref{eqn:r2} using $\beta=0.42$ for PS/10CB and $\beta=0.40$ for PS/12CB). \label{fig:comp}}
\end{figure}

Based upon the approach of Benmouna et al \cite{Benmouna1999}, the DSC data (shown in \ref{fig:dsc}) was utilized to estimate the solubility limit of the liquid crystal in the polymer $\beta$ (defined as the maximum amount of liquid crystal that can be dissolved in the polymer at $T_{AI}$, expressed as a weight fraction) and the relative amount of liquid crystal in the liquid crystal-rich domains $\alpha$ (equal to the ratio of the weights of liquid crystal in the liquid crystal-rich domains over the total liquid crystal). The solubility limit $\beta$ can be determined from the change of enthalpy at the isotropic/smectic-A transition for the smectic PDLC material and that of the pure liquid crystal \cite{Benmouna1999}:
\begin{equation} \label{eqn:delta1}
\delta =  \frac{\Delta H_{AI}^{PDLC}}{\Delta H_{AI}^{LC}}
\end{equation}
where $\delta$ is the ratio of the enthalpies, $H_{AI}^{PDLC}$ is the enthalpy of phase-ordering of the PDLC material, and $H_{AI}^{LC}$ is the enthalpy of phase-ordering of the pure liquid crystal. This ratio $\delta$ can be related to the solubility limit $\beta$ using the rule of inverse segments \cite{Benmouna1999}:
\begin{equation}\label{eqn:delta2}
\delta =  \frac{w_{LC} - \beta}{1-\beta}
\end{equation}
where $w_{LC}$ is the weight fraction of liquid crystal in the PDLC material. This approach is valid given the assumptions \cite{Benmouna1999} that the thermodynamic properties of the liquid crystal in both domains (liquid crystal- and polymer-rich) are the same, the amount of liquid crystal dissolved in the polymer is constant for all materials where $w_{LC} \ge \beta$, and the isotropic/smectic-A transition temperature is not dependent on $w_{LC}$ (well supported by the observations presented in \ref{fig:exp_phase_diag}a,b). \ref{fig:comp}c,d show $\delta$ versus $w_{LC}$ for both smectic PDLC materials (determined using eqn. \ref{eqn:delta1}), where extrapolation to $\delta = 0$ yields the solubility limit $\beta$ (at a reference temperature of $T_{AI}$ from \ref{fig:exp_phase_diag}a,b). The values of $\beta=0.40$ for PS/12CB and $\beta = 0.42$ for PS/10CB were determined. The fraction of liquid crystal in both polymer- and liquid crystal-rich domains can also be calculated from $\delta$ \cite{Benmouna1999}, under the assumption that the liquid crystal-rich phase is essentially pure:
\begin{equation}\label{eqn:r1}
r =  \frac{\delta}{\phi_{LC}}
\end{equation}
where $r$ is the mass fraction liquid crystal in the LC-rich phase over the total liquid crystal mass in the material. This can also be calculated again using the rule of inverse segments \cite{Benmouna1999}:
\begin{equation}\label{eqn:r2}
r =  \frac{\phi_{LC} - \beta}{\phi_{LC} \left( 1-\beta\right)}
\end{equation}
\ref{fig:comp}e,f show $r$ versus $\phi_{LC}$ for both smectic PDLC material (determined using eqn. \ref{eqn:r1}) qualitatively agreeing well with past results \cite{Benmouna1999} where $r$ approaches zero as the solubility limit is approached (and no LC-rich phase exists) and reaches one as the polymer content approaches zero.

\subsection{Macro-scale Morphologies}

\begin{figure*}[htp]
\centering
\subfigure[]{\includegraphics[angle=90,width=0.25\linewidth]{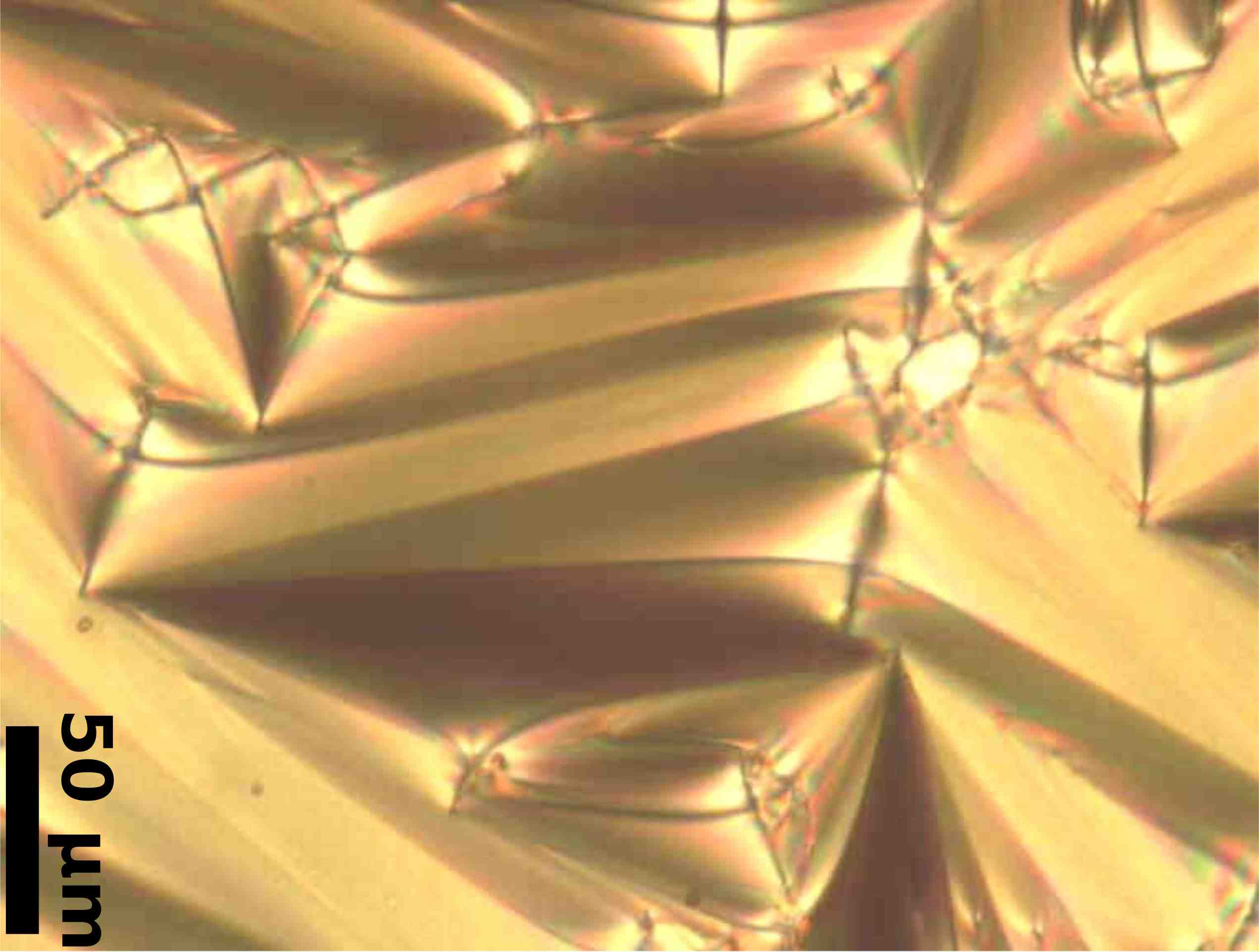}}\subfigure[]{\includegraphics[angle=90,width=0.25\linewidth]{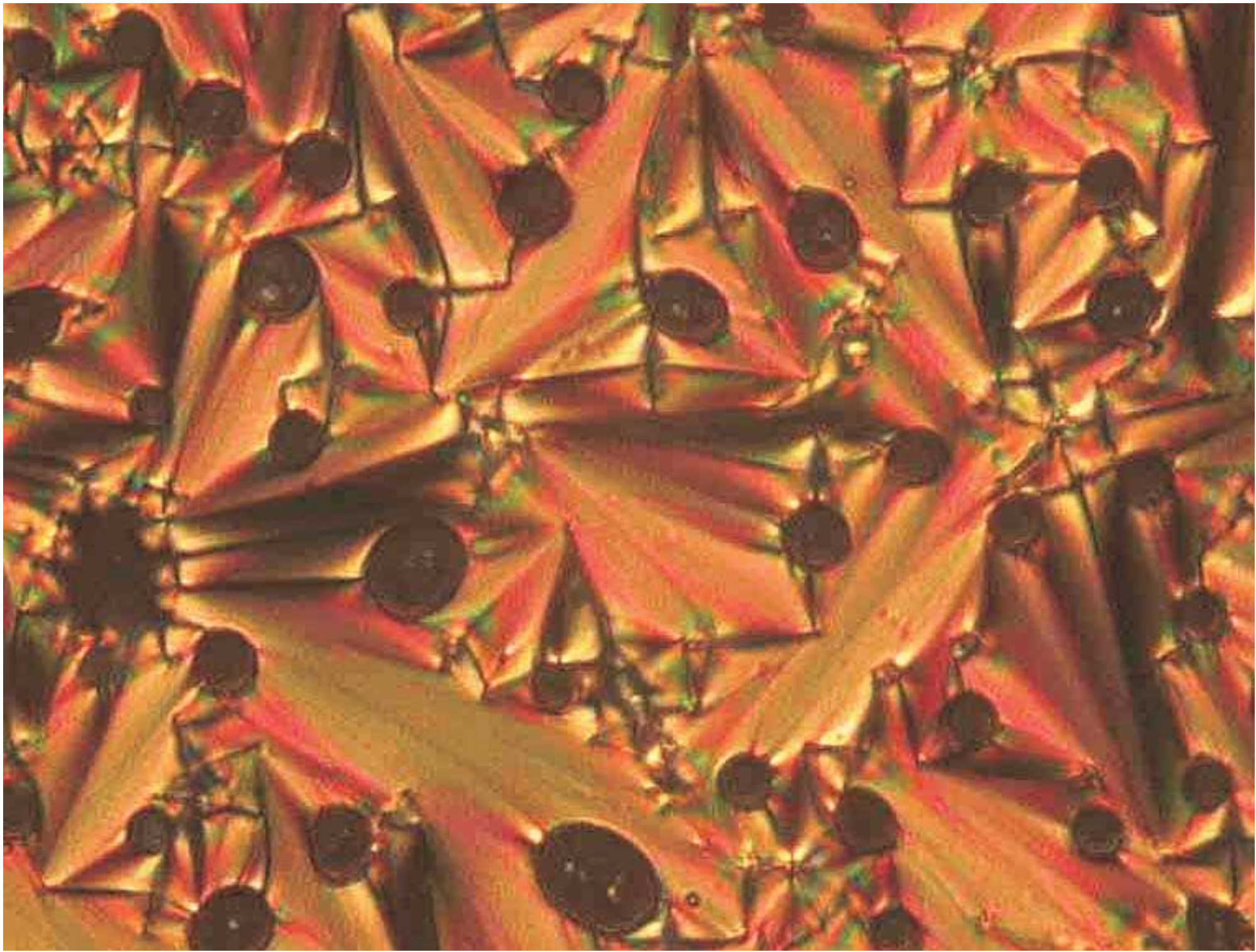}}\subfigure[]{\includegraphics[angle=90,width=0.25\linewidth]{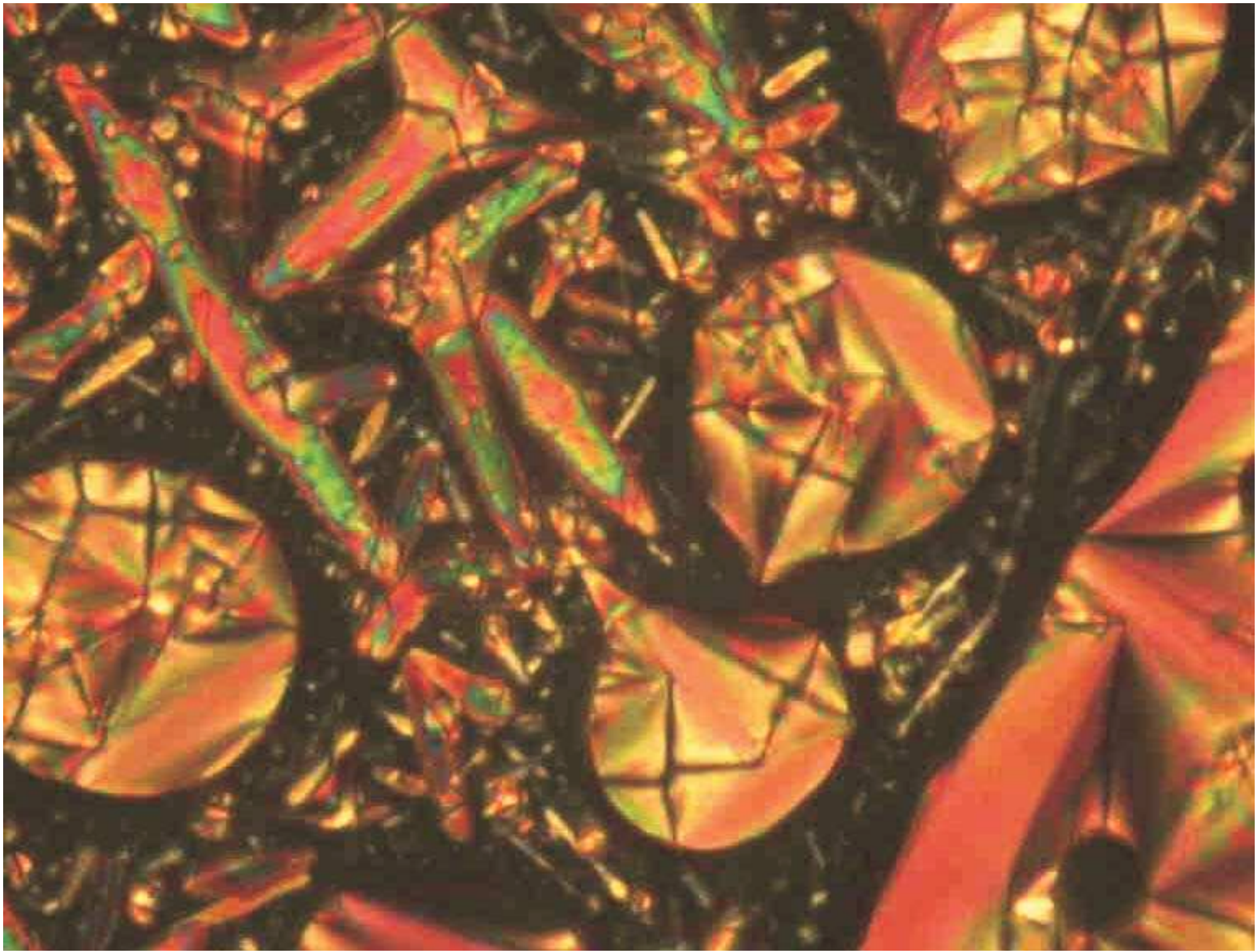}}\subfigure[]{\includegraphics[angle=90,width=0.25\linewidth]{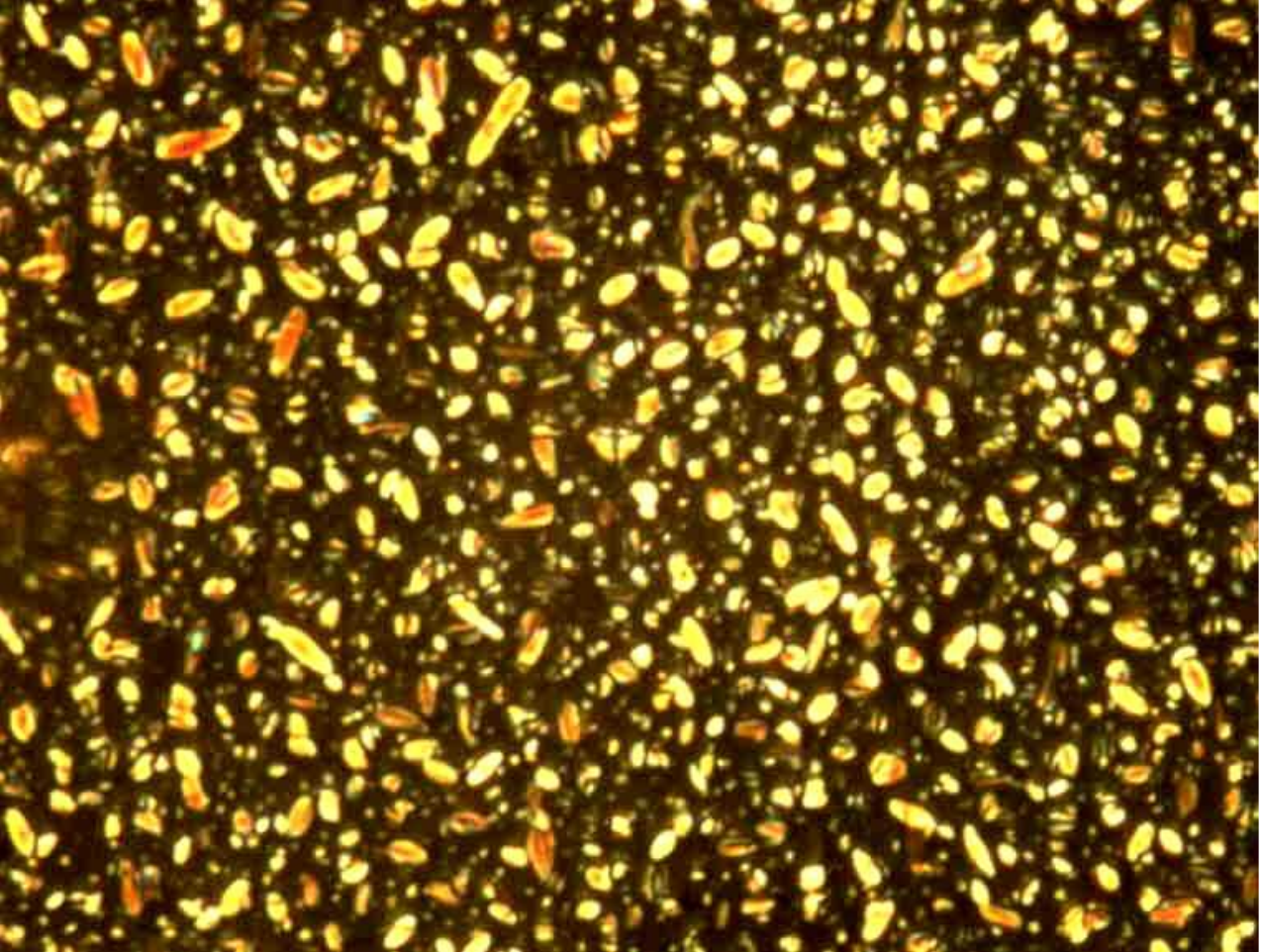}}
\caption{POM images of LC/PDLC morphologies obtained by cooling from $52^\circ C$ to $50^\circ C$ at $0.5^\circ C/min$: (a) $100.0 \%wt$ 10CB, (b) $95.5\%wt$ 10CB, (c) $75\%wt$ 10CB, (d) $49.7\%wt$ 10CB.
 \label{fig:morpho10}}
\end{figure*}

The determined phase diagrams (\ref{fig:exp_phase_diag}) provide qualitative guidance on which phase will be continuous and which will be dispersed, or if both will be co-continuous (depending if the concentration is lower, higher or very close to the critical concentration), but is not sufficient to make quantitative predictions on the actual morphologies. In the case of PDLCs, the presence of both phase separation and phase ordering transitions results in several different dynamic pathways to affect domain morphology of the final material. Three main categories of factors are involved: phase transition sequence, phase transition kinetics, and the introduction of anisotropy/texture from the presence of the liquid crystal. Phase transition sequence, in the context of the experimentally observed equilibrium phase behavior of smectic PDLCs (\ref{fig:exp_phase_diag}), involves either a staggered initiation of phase separation/phase ordering (when $T_{II}$ and $T_{AI}$ exist) or a simultaneous initiation (when $T_{I/AI}$ exists, or when the sample is rapidly cooled such that I/I phase separation is negligible). The presence of anisotropy in these materials, as mentioned before, introduces an additional degree of complexity. 

A unique attribute to liquid crystals exhibiting a direct disordered/ordered smectic transition are the complex set of growth morphologies observed, which have been the focus of much past study \cite{Rey2007} (Figure \ref{fig:morpho10}c). The smectic-A mesophase, with its lamellar ordering on the molecular scale, exhibits growth, defect, and texture phenomena not seen in the nematic phases, including the lamellar-like cholesteric mesophase. In addition to spheroids and sphero-cylinders observed in nematics, direct isotropic/smectic-A liquid crystals exhibit the fascinating highly-anisotropic ``batonnet'' structures of Friedel and Grandjean \cite{Friedel1910,Dierking2003,Dierking2003a,Abukhdeir2009} and complex buckling filaments \cite{Naito1997,Weinan1999,Todorokihara2001,Todorokihara2004a,Rey2008a}. As there are many factors involved, a discussion of the phase transformation processes in these materials will be presented before analyzing the experimental results in detail, in order to guide the interpretation of the observed morphologies.

For a smectic PDLC material where both $T_{II}$ and $T_{AI}$ exist, quenching from the fully isotropic one-phase state below $T_{II}$ results in phase separation occurring into fully isotropic liquid crystal-rich and polymer-rich domains. These domains formed during phase separation into two different isotropic phases result in what will be referred to as ``stage I'' PDLC morphology. In this stage spherical domains are formed, unaffected by whether the polymer-rich or liquid crystal-rich phase is the matrix phase, in that both phases are isotropic. This process is well understood as a standard liquid/liquid demixing process where the size and number density of the spherical domains formed are determined by liquid/liquid interactions and quench rate \cite{Cahn1965,Chan1996,Chan1997}. As domains formed in stage I are quenched below $T_{AI}$, their morphology is typically unaffected by the phase-ordering transition and standard smectic transition kinetics and coarsening are observed \cite{Abukhdeir2008a} (within the liquid crystal-rich domain). In the case where the liquid crystal-rich phase is the matrix phase (concentration of LC higher than the critical concentration, see \ref{fig:morpho10}b), the spherical polymer-rich domains serve to prohibit the formation of a neat liquid crystalline texture and the final texture can be predicted given that the liquid crystal/polymer interface preferred anchoring is known \cite{Gupta2005,Das2005a,Das2006}. 


Two different phase transition sequences can result in stage I spherical morphologies: a) quenching a fully isotropic PDLC below $T_{II}$ held for long times (complete phase separation) and b) quenching a fully isotropic PDLC below $T_{II}$ held for short times and then quenching again below $T_{AI}$. When the material is cooled below $T_{AI}$, the liquid crystal diffuses from already formed isotropic polymer-rich domains to neighboring liquid crystal-rich domains, but if this diffusion is slow compared to the cooling rate, the isotropic polymer-rich phase becomes super-saturated in the LC component. This results in secondary phase separation, with formation of new domains, which are referred to as ``stage II,'' and include both newly formed domains (post-stage I) and small domains formed during stage I (lengths on the order of the characteristic length of liquid crystalline ordering). Phase separation in this regime results in a myriad of complex morphologies in that surface tension anisotropy is present, in addition to the surface elastic effects of bulk liquid crystalline texture. Much experimental research has been done on characterizing growth morphologies of both pure and lyotropic liquid crystals which exhibit a direct isotropic/smectic-A transition \cite{Fournier1991,Pratibha1992,Pratibha1995,Naito1997,Todorokihara2001,Todorokihara2004a}. Much of the interesting morphologies exhibited, including spherical, sphero-cylindrical, and filament domains \cite{Naito1997,Todorokihara2001,Todorokihara2004a}, are attributed to the presence of interfacial tension anisotropy $\Delta \gamma = \gamma_\parallel - \gamma_\perp$ which is coupled to the concentration of solvent present (or polymer in the case of PDLCs) and degree of liquid crystallinity.

The final type of morphological growth occurs at longer time-scales than stage I-II growth, where liquid crystal-rich spherulites in the micron/sub-micron range form in polymer-rich domains that are approaching equilibrium concentration. In this case, diffusion of the liquid crystal is increasingly less energetic, becoming the kinetically controlling step and limiting the liquid crystal-rich domain size that can be locally formed. As diffusion becomes predominant in this regime, the domains formed are not only smaller, but also more symmetric than the stage II domains. This stage is favored at high polymer concentration, where the molecular mobility is lower.

For PDLC materials with polymer concentrations higher than the maximum solubility limit (as defined previously) both phase separation and ordering occur simultaneous and $T_{I/AI}$ is found (refer to \ref{fig:exp_phase_diag}). In this case, stage I morphologies are not observed in that isotropic/isotropic phase separation does not take place. Thus depending on the relative ratio of polymer to liquid crystal, stage II morphologies are observed (favored in mixtures with relatively less polymer) in addition to stage III morphologies (favored in mixtures with relatively more polymer). PDLCs in these concentration ranges are more difficult to study experimentally because of the small domain sizes formed (order of $\mu m$). 

A simple scaling argument can be derived to describe the characteristic micron size of a drop formed in stage III. The simplest LC shape equation is a balance of the bulk stress jump and anisotropic surface tension \cite{Rey2007}:
\begin{equation} \label{eqn:shape_eq}
- \left( \bf{k \cdot T}_b \right) \cdot \bf{k} =  \left[ \gamma \bf{I}_s + \frac{\partial^2\gamma}{\partial\bf{k}^2}\right]:\bf{e}_1
\end{equation}
where $-\left( \bf{k \cdot T}_b \right) \cdot \bf{k}$ is the stress jump, $\bf{k}$ is the unit normal, $\gamma$ is the interfacial tension, H the average curvature, D the deviatoric curvature , $\bf{I}_s=\bf{e}_1 \bf{e}_1 + \bf{e}_2 \bf{e}_2$ is the unit surface tensor, $\bf{q}=\bf{e}_1 \bf{e}_1 - \bf{e}_2 \bf{e}_2$ is the deviatoric unit tensor, and $\bf{e}_i$ are the principal eigenvectors of the curvature tensor $\bf{e}_1$ . For spherical shapes ( $D=0, H=-1/R$ ) and neglecting anchoring effects ($\frac{\partial^2\gamma}{\partial\bf{k}^2}:\bf{I}=0$) the drop radius $R$ is
\begin{equation} \label{eqn:radius_1}
R=\frac{2\gamma}{ \left( \bf{k \cdot T}_b \right) \cdot \bf{k}}
\end{equation}

The interfacial tension $\gamma$ and bulk stress jump $\left( \bf{k \cdot T}_b \right) \cdot \bf{k}$ can be expressed in terms of the homogeneous free energy density $f_h$:
\begin{equation} \label{eqn:gamma}
\gamma=2\sqrt{L_{\Psi}}\int_{\Psi_e}^0 \! \left[ f_h(\Psi,c)-f_h(\Psi_e,c_e) \right]^{1/2}  \, d\Psi
\end{equation}
where the subscript $e$ denotes the bulk equilibrium value, and $L_{\Psi}$ is the gradient elasticity for the smectic order parameter. Replacing these expressions into equation \ref {eqn:radius_1}, we find that the spherulite radius $R$ is given by a balance between the bulk energy with the gradient energy:
\begin{equation} \label{eqn:radius_2}
R= C \sqrt{\frac{L_{\Psi}}{kT|c^*-c_e|}} \, ; \, C=\frac{\int_{\Psi_e}^0 \! 2\left[ \bar{f}_h(\Psi,c)-\bar{f}_h(\Psi_e,c_e) \right]^{1/2}  \, d\Psi}{\bar{f}_h(\Psi_e,c_e)}
\end{equation}
where the scaled free energy is $\bar{f}_h=f_h/kT|c^*-c_e|$ and $c^*$ is the concentration at which the isotropic phase is unstable. Using $C=1$, we find that at coexistence $R$ is given by :
\begin{equation} \label{eqn:radius_3}
R \sim \sqrt{\frac{L_{\Psi}}{kT|c^*-c_e|}} 
\end{equation}
Assuming $L_{\Psi}\approx 1 pN$ and $kT|c^*-c_e| \approx 1 \frac{J}{m^3}$ eqn. \ref{eqn:radius_3} provides an estimate of $R\approx 1 \mu m$. Once the droplet forms, the growth process is controlled by diffusion. Under diffusion-limited conditions, further growth is inhibited and the spherulites remain in the low $\mu m$ range (see \ref{fig:morpho10}d and \ref{fig:morpho12}d).


\subsubsection{Quasi-equilibrium morphologies: concentration effects}

\begin{figure*}[htp]
\centering
\subfigure[]{\includegraphics[angle=90,width=0.25\linewidth]{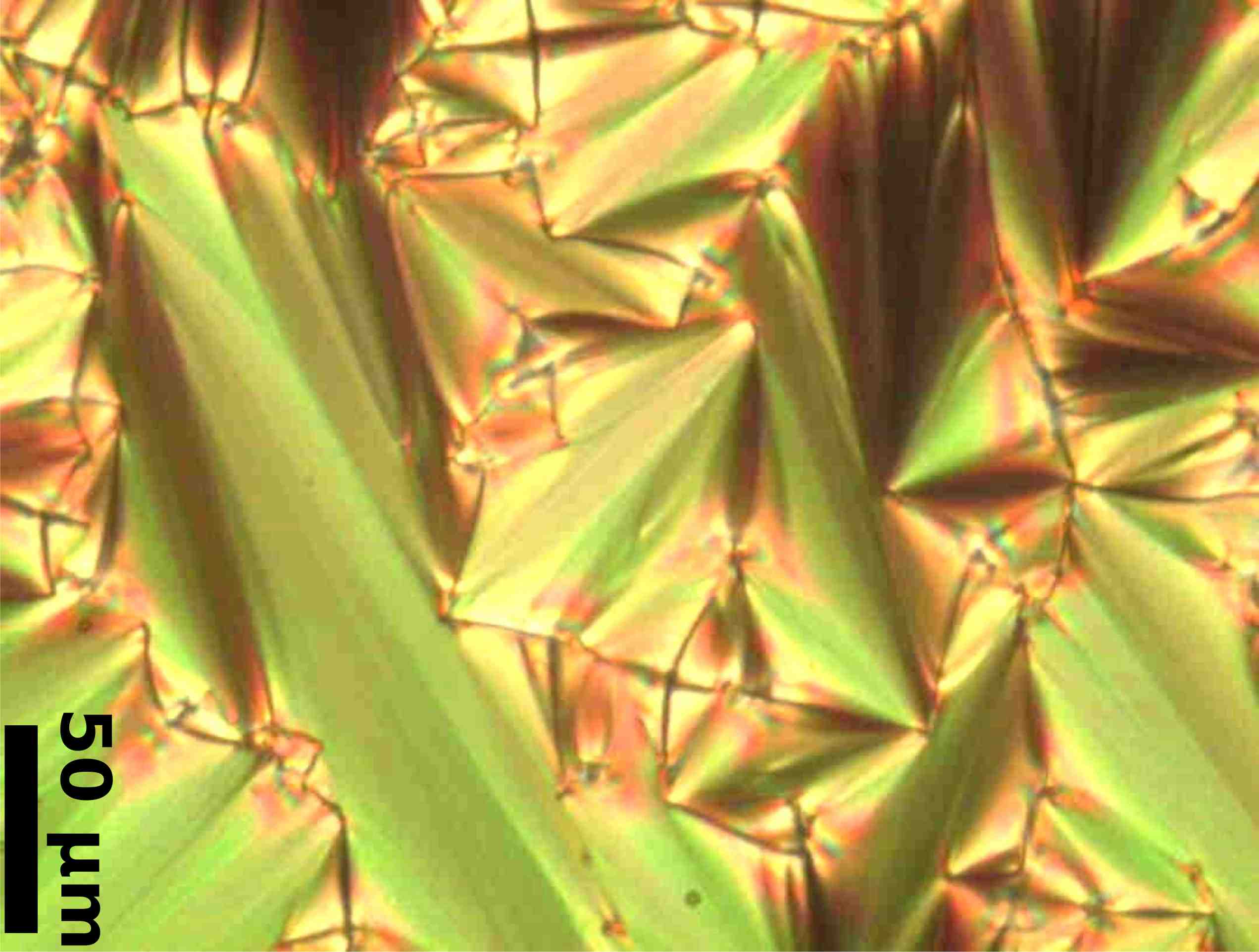}}\subfigure[]{\includegraphics[angle=90,width=0.25\linewidth]{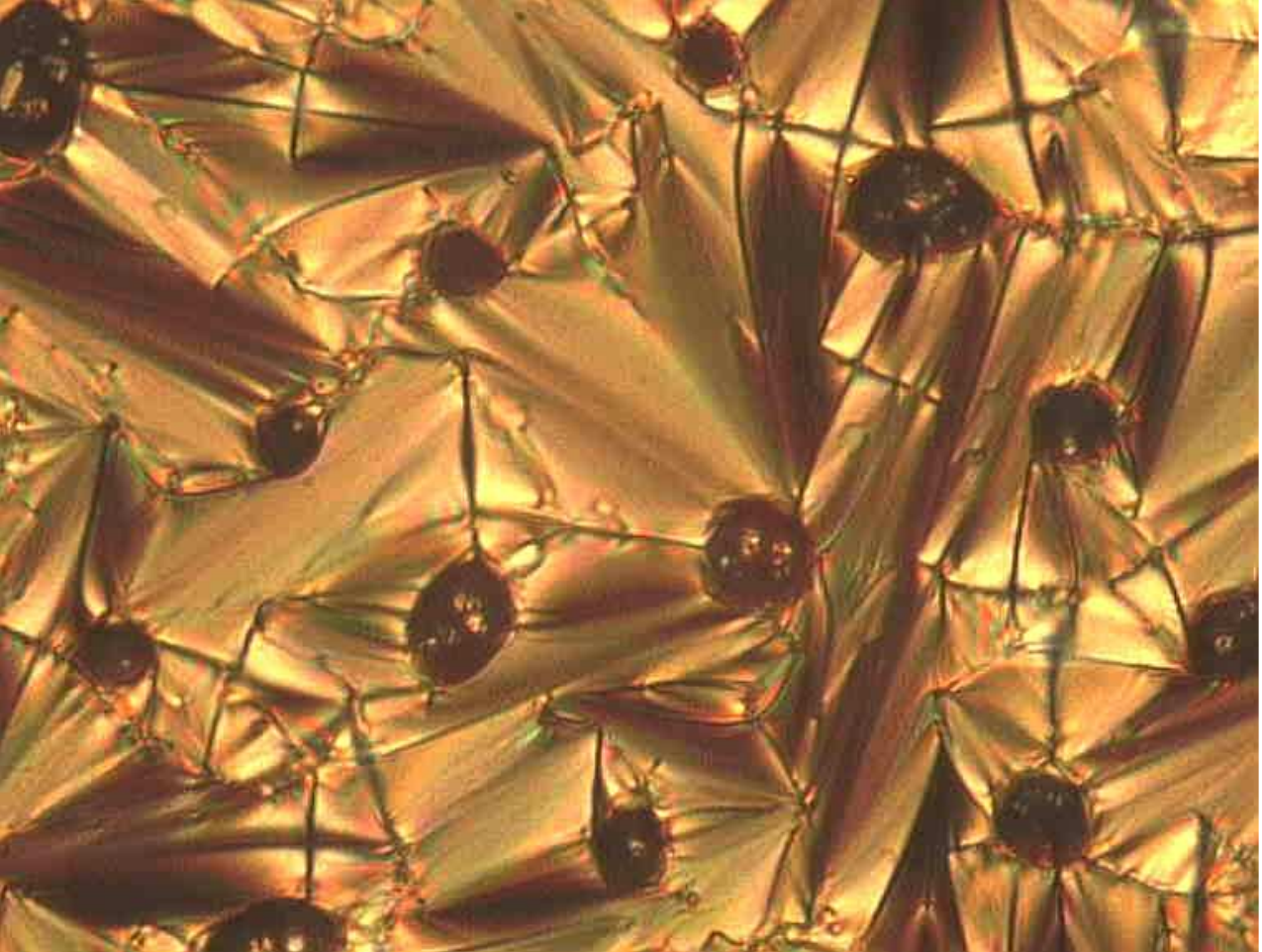}}\subfigure[]{\includegraphics[angle=90,width=0.25\linewidth]{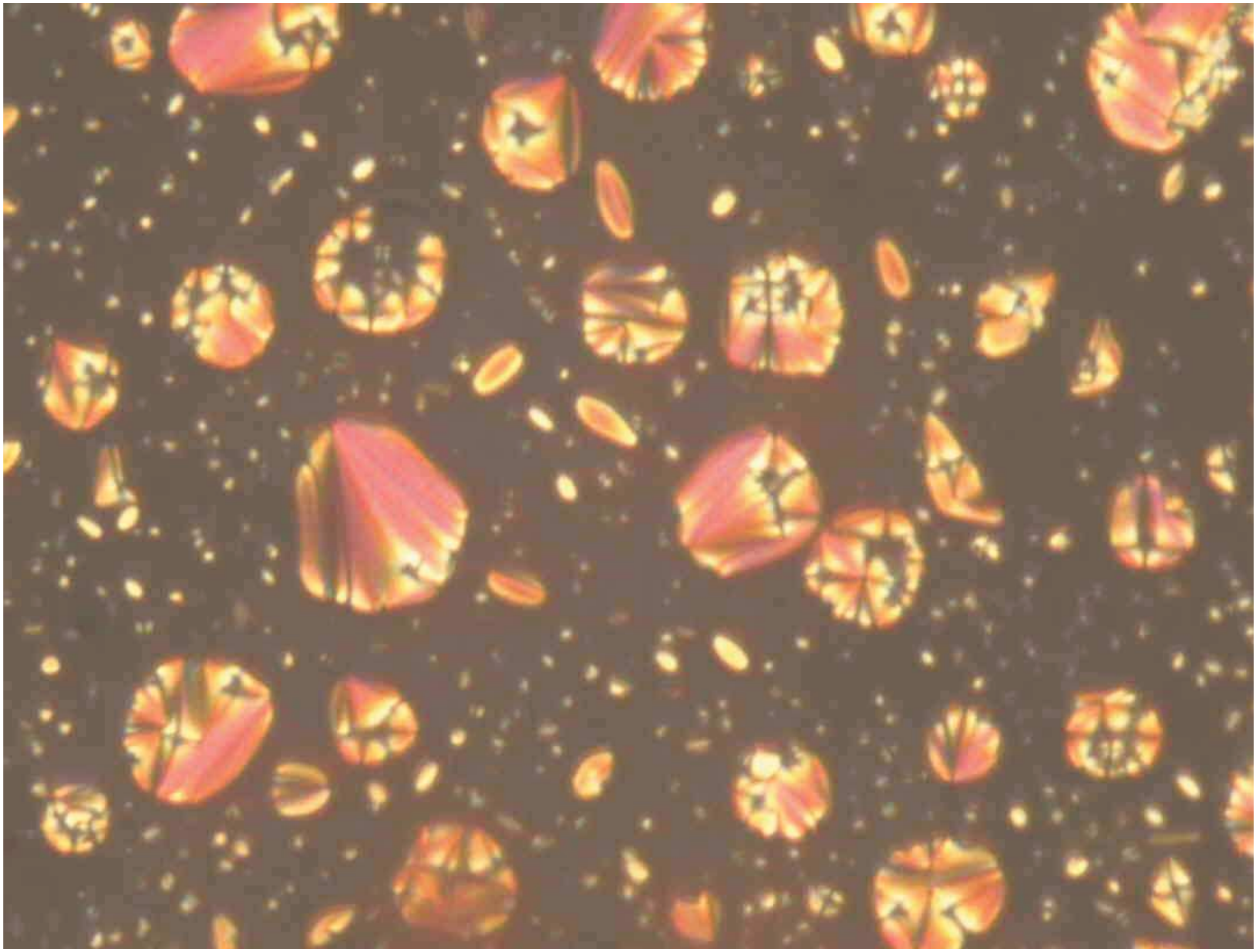}}\subfigure[]{\includegraphics[angle=90,width=0.25\linewidth]{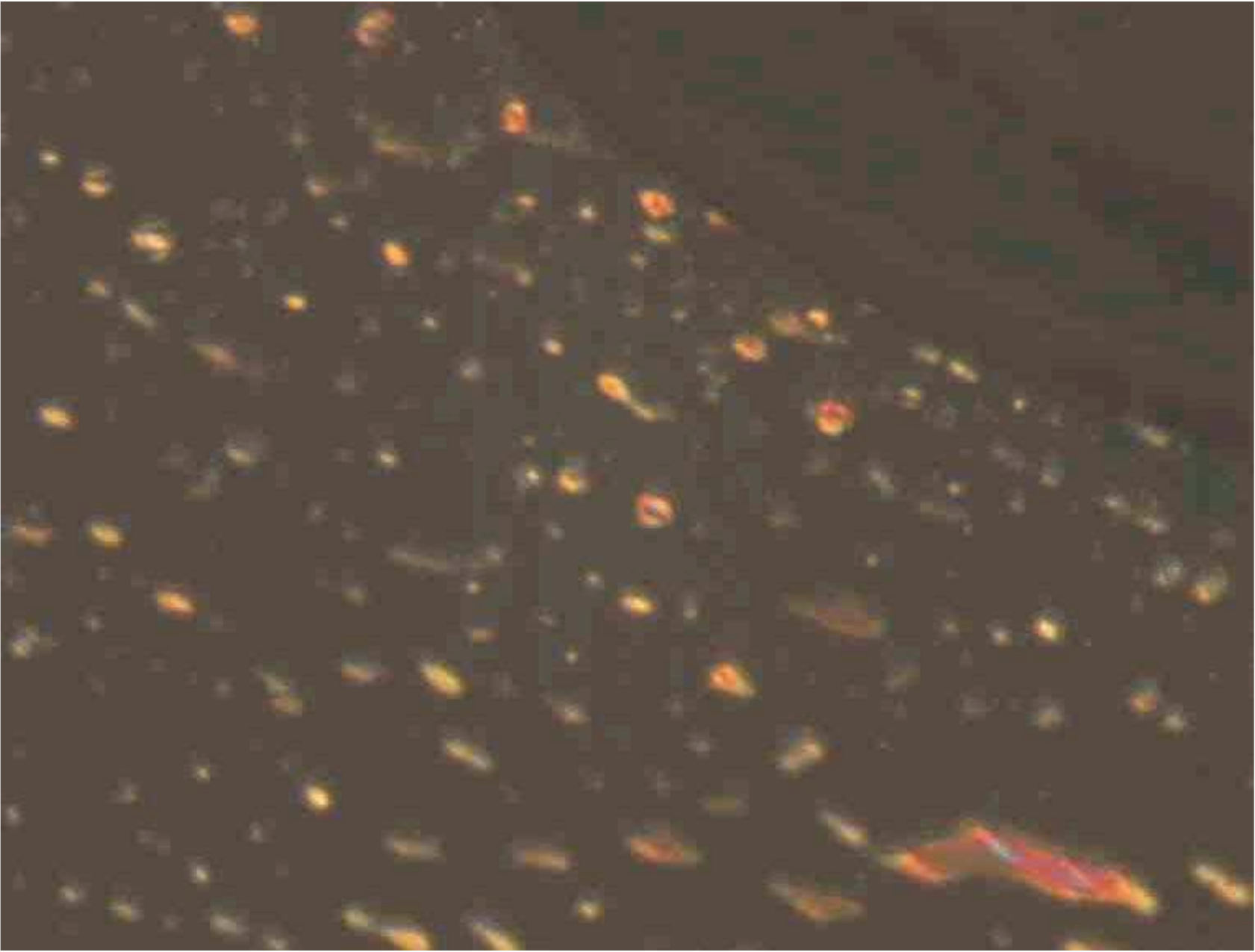}}
\caption{POM images of LC/PDLC morphologies obtained by cooling from $60^\circ C$ to $55^\circ C$ at $0.5^\circ C/min$. (a) $100\%wt$ 12CB, (b) $95.3\%wt$ 12CB, (c) $62.5\%wt$ 12CB, (d) $30\%wt$ 12CB. \label{fig:morpho12}}
\end{figure*}

\ref{fig:morpho10}and \ref{fig:morpho12} show representative quasi-equilibrium morphologies obtained for mixtures with different concentrations of 10CB and 12CB, respectively. The samples were heated from room temperature to a temperature above the isotropic/smectic-A transition (60$^\circ$C for mixtures with 12CB and 52$^\circ$C for mixtures with 10CB), and then they were cooled at 0.5$^\circ$C/min and held at the final temperature for $20$ minutes before the images were taken. The initial temperatures were below the isotropic/isotropic solubilization temperature for mixtures with more than $50\%wt$ liquid crystal, such that some degree of phase separation took place before quench below the phase-ordering transition. \ref{fig:morpho10}a and \ref{fig:morpho12}a show the morphologies of the pure liquid crystal, where the typical focal conic texture of a smectic phase can be observed \cite{Dierking2003a}. As polymer concentration is increased, a two-phase morphology can be seen. \ref{fig:morpho10}b and \ref{fig:morpho12}b show the morphologies of samples with 95\%wt LC (above the the critical-point concentration, which was calculated to be 0.87 for 10CB and 0.88 for 12CB) where the liquid crystal-rich phase is the matrix. Isotropic/isotropic phase separation produces a droplet morphology (stage I) where the spherical dispersed polymer-rich domains form in the liquid crystal-rich matrix. Once quenched below $T_{AI}$, the continuous liquid crystal-rich matrix phase transitions to smectic-A via the standard nucleation/growth mechanism with additional topological constraints imposed by the dispersed polymer-rich domains (see previous discussion). Additionally, small liquid crystal-rich domains are observed to form inside the isotropic polymer-rich droplets, giving rise to a ``salami''-type structure frequently observed in polymer mixtures (for example, high-impact polystyrene \cite{Wagner1970,Echte1977}). The formation of these morphologies is due to secondary phase separation, as explained before (stages II-III).

When the concentration of liquid crystal becomes lower than the critical concentration, the isotropic phase is the matrix and the smectic phase is dispersed. The size of the smectic domains becomes smaller as the concentration of liquid crystal decreases. In this sub-critical concentration range of liquid crystal, the morphology of the material is composed of a combination of large and small domains, where the large domains are predominantly spherical in shape and the small domains have a much more asymmetric shape. The spherical domains were formed by the I/I phase separation (stage I), and the smaller and asymmetric domains correspond to the stages II-III. In \ref{fig:morpho10}d and \ref{fig:morpho12}d, the mixture composition is such that there is no I/I equilibrium, so stage I morphologies are not observed.

\subsubsection{Effects of quench rate on morphology dynamics}

\begin{figure*}[htp]
\centering
\subfigure[]{\includegraphics[width=0.30\linewidth]{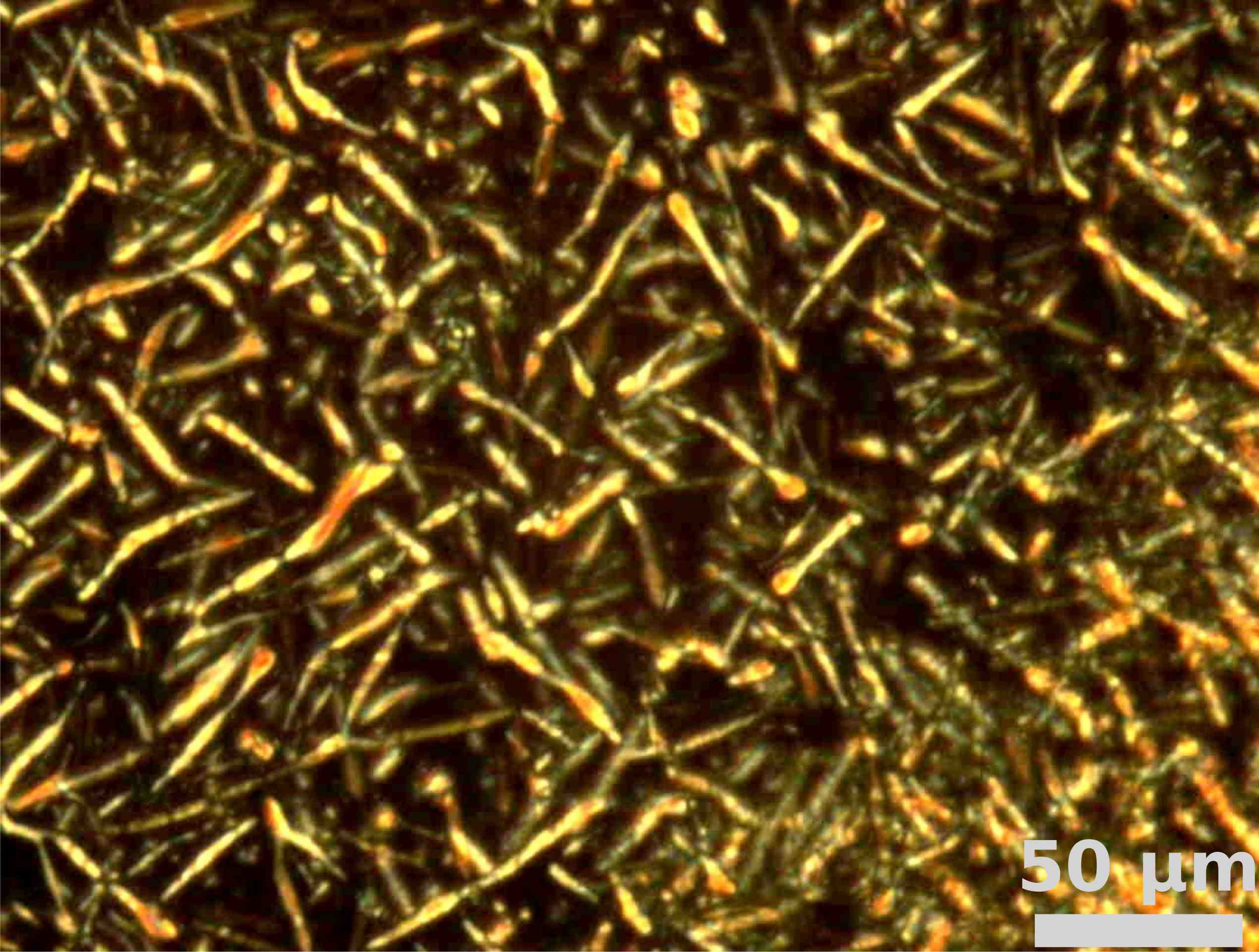}}\subfigure[]{\includegraphics[width=0.30\linewidth]{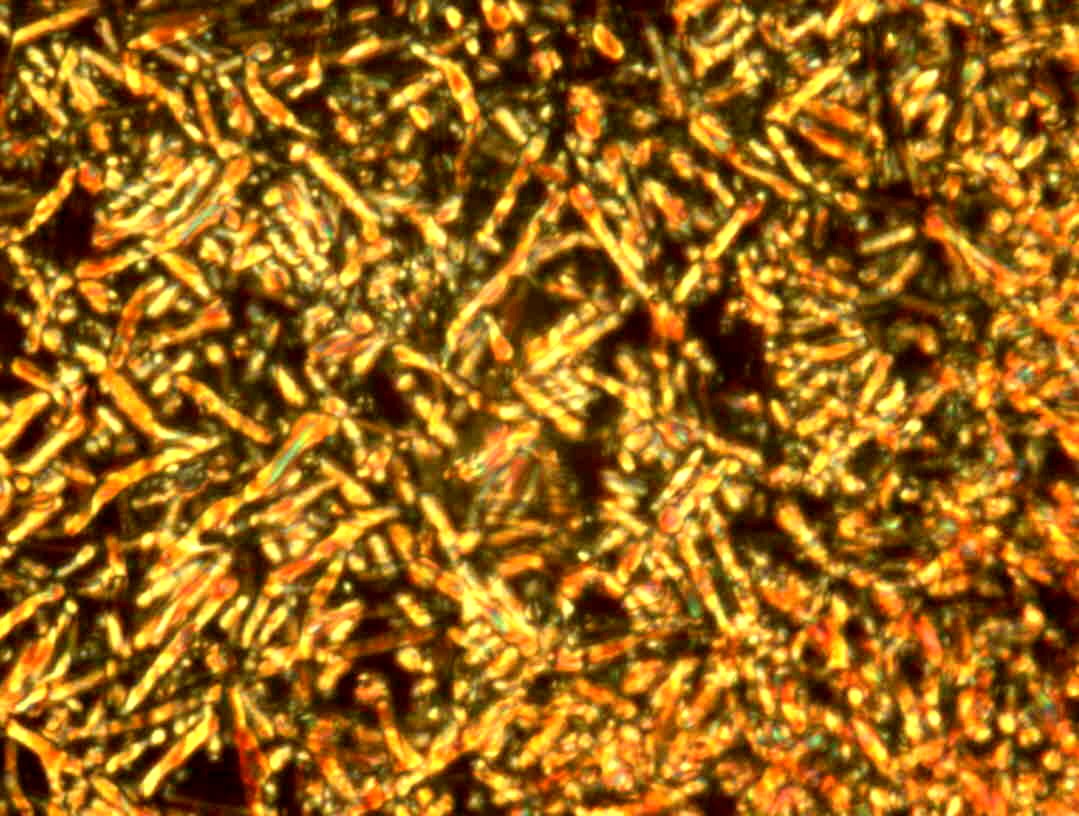}}\subfigure[]{\includegraphics[width=0.30\linewidth]{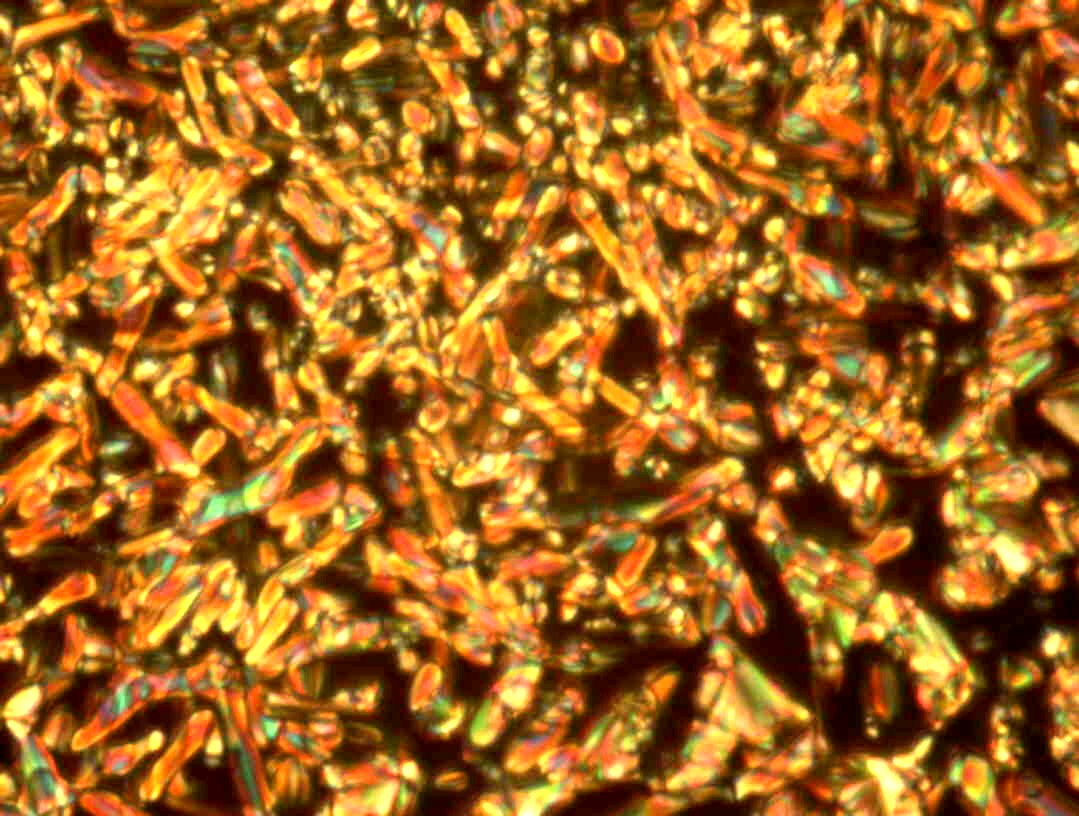}}\\
\subfigure[]{\includegraphics[width=0.30\linewidth]{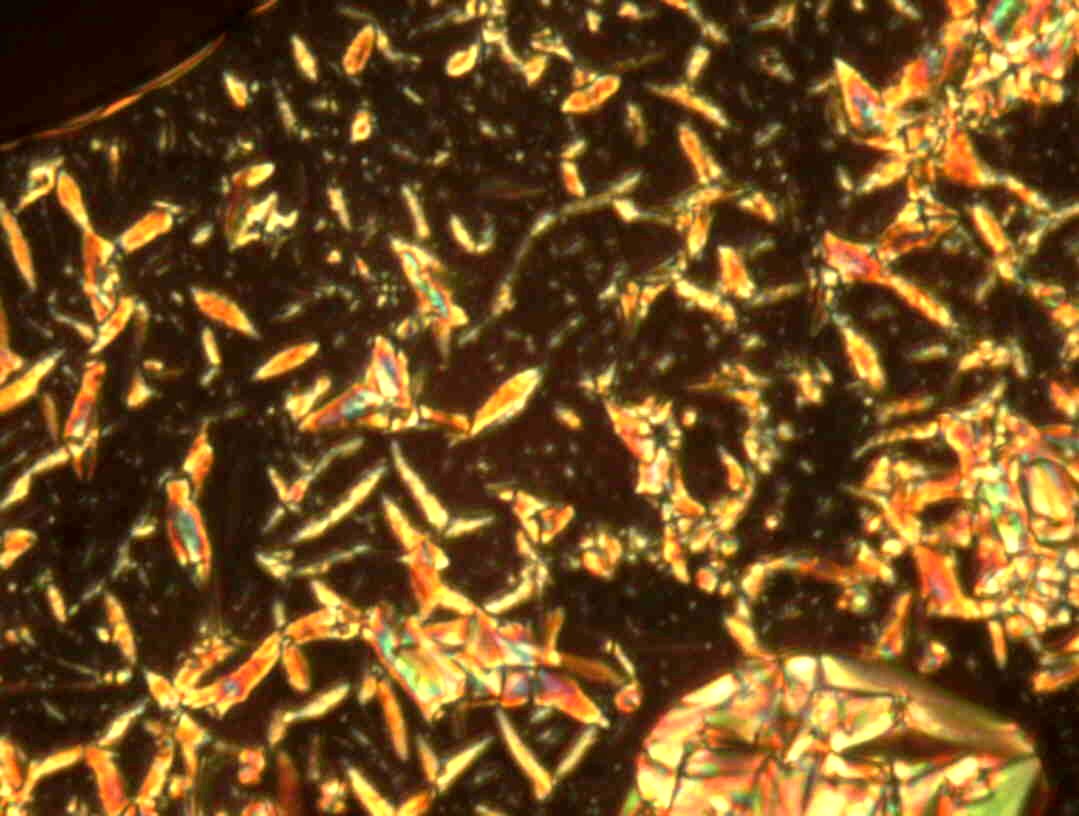}}\subfigure[]{\includegraphics[width=0.30\linewidth]{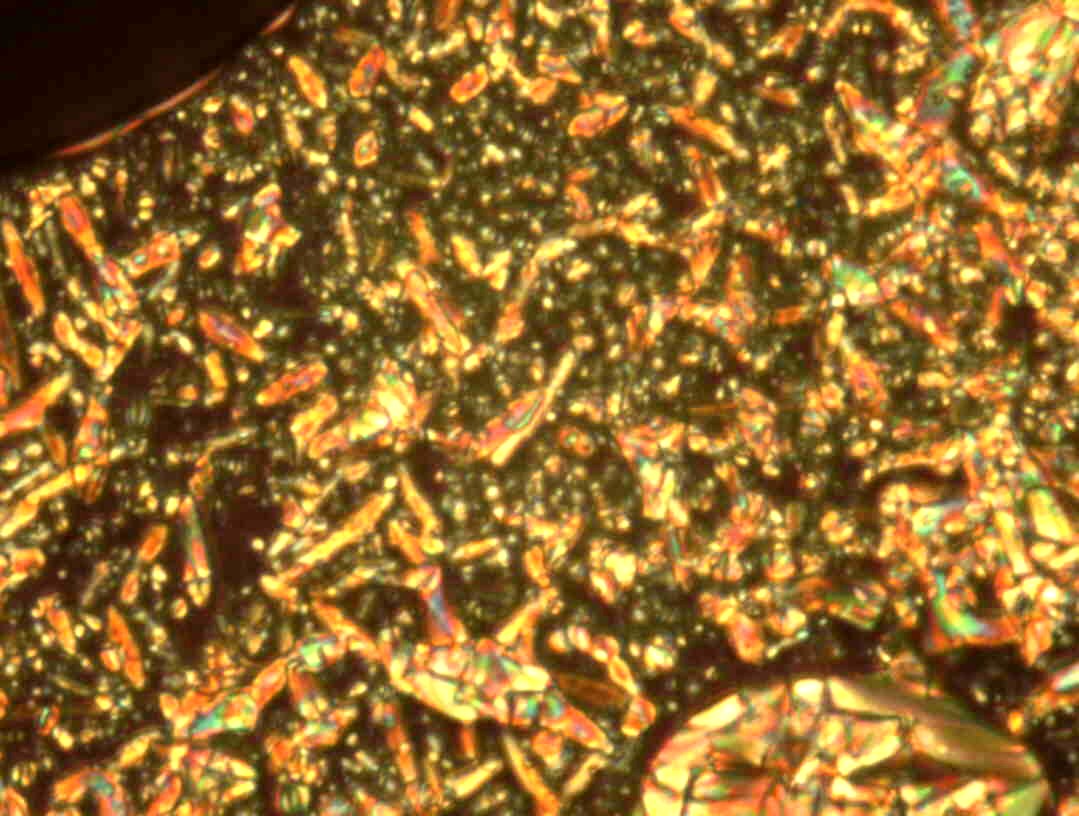}}\subfigure[]{\includegraphics[width=0.30\linewidth]{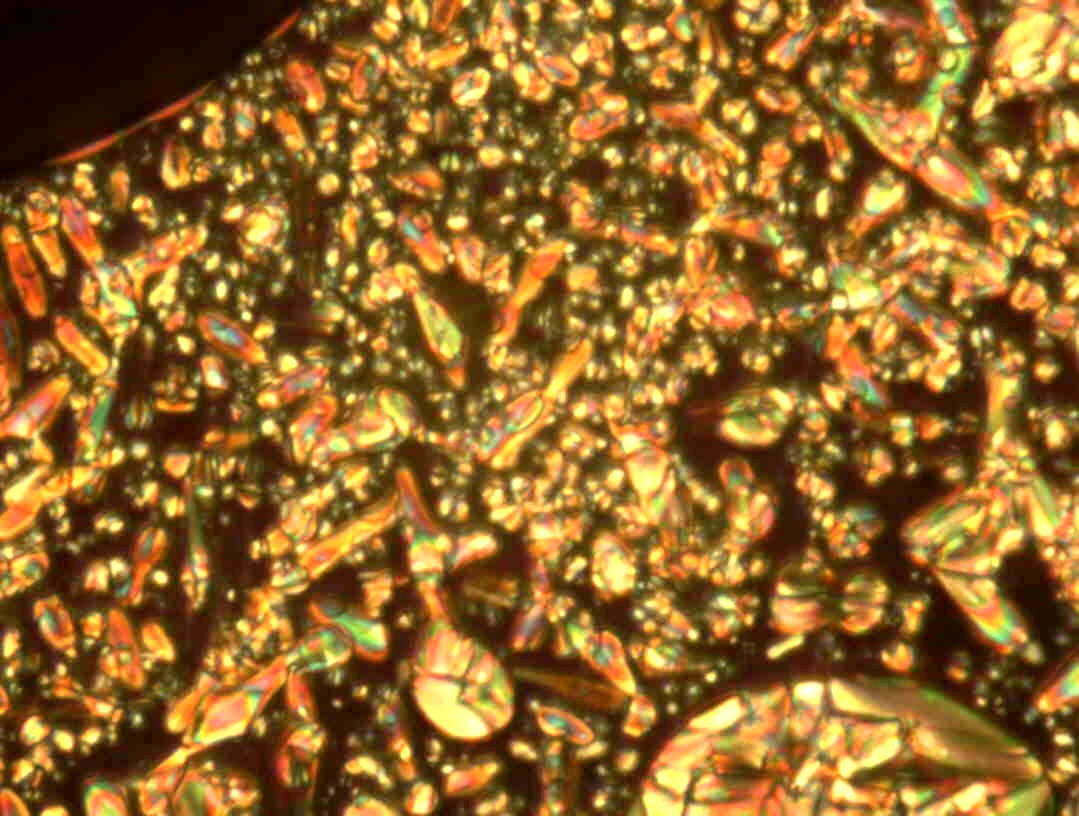}}
\caption{POM images of LC/PDLC morphologies obtained after quenching a mixture with 85.1\% 12CB. starting at 80$^\circ$C, quench at 20$^\circ$C/min to 50$^\circ$C. the snapshots were taken at (a) 1:25, (b) 2:45, and (c) 10:00 minutes. Quench starting at 80$^\circ$C, quench at 20$^\circ$C/min to 53$^\circ$C, hold for 10 minutes, quench at 20$^\circ$C/min to 50$^\circ$C. The snapshots were taken at (d) 0:14, (e) 1:46, and (f) 10:00 minutes
 \label{fig:quench_85_12CB}}
\end{figure*}

In order to further study the formation and dynamics of different morphologies, two types of rapid quenches from a temperature above $T_{II}$ were performed: (a) direct quench at 20$^\circ$C/min to the isotropic/smectic-A coexistence region and (b) quench at 20$^\circ$C/min to the isotropic/isotropic region, hold for 10 minutes, quench at 20$^\circ$C/min to the isotropic/smectic-A region. In quench (a) isotropic/isotropic phase separation is minimized such that stage I morphologies are rare, whereas in quench (b) isotropic/isotropic phase-separation is allowed to progress, resulting in many spherical stage I domains. \ref{fig:quench_85_12CB} shows the time-progression of the domain morphologies at different times during the quench sequence (a) (directly into the isotropic/smectic-A regime), for a mixture containing 85\%wt of 12CB. It is observed that, in this case, the vast majority of smectic-A domains formed correspond to stage II morphology. \ref{fig:quench_85_12CB} shows the time-progression of quench (b) where distinct stages I-III are formed, directly exhibiting the influence of isotropic/isotropic phase separation preceding phase-ordering. In the direct quench, when the phase separation/transition process begins the matrix has a relatively high concentration of liquid crystal, such that a large majority of stage II domains form. In the two-stage quench, due to isotropic/isotropic phase-separation, secondary phase separation/transition takes place after the final quench in a matrix phase with a relatively high concentration of polymer, such that stage III morphologies are prevalent also.

\subsection{Nano-scale Structure} \label{sec:simulation}

\begin{figure*}[htp]
\centering
\subfigure[]{\includegraphics[width=0.3\linewidth]{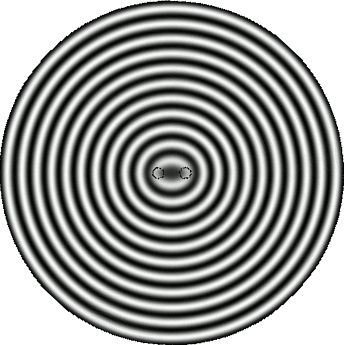}}
\subfigure[]{\includegraphics[width=0.3\linewidth]{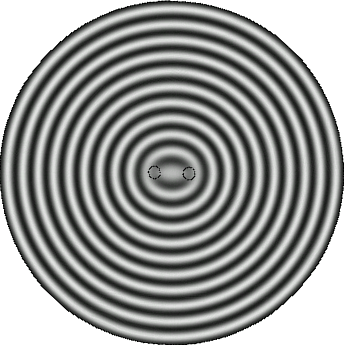}}
\subfigure[]{\includegraphics[width=0.3\linewidth]{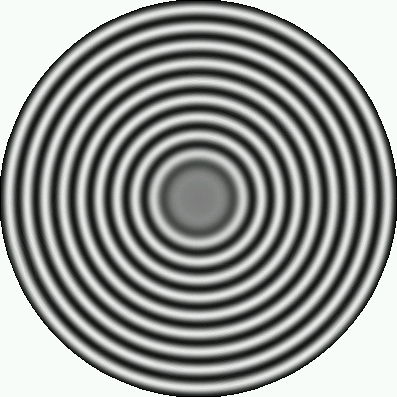}}
\caption{Simulation results of the smectic density modulation $Re(\Psi)$ (surface) and biaxiality (contour) $\beta^2$: (a) $T=331.00 K$, (b) $T=331.35 K$, and (c) $T=331.41 K$. Surface plot values range from $Re(\Psi)=-0.161$ (black) to $Re(\Psi)=0.161$ (white) and the contour corresponds to $\beta^2=0.1$. \label{fig:struct1}}
\end{figure*}

In the final stage of the study, the material transformation model was simulated in a two-dimensional circular smectic-A domain. Initial and boundary conditions (see Sections \ref{sec:mattrans_model}) were used to simulate a bulk radial-oriented smectic-A domain, similar to past studies \cite{Kralj1996,Rey2008a}. Previously determined phenomenological parameters were used based on 12CB \cite{Abukhdeir2008c} to perform a parametric study in temperature, with fixed cylindrical length scale of $97.5 nm$. Time-dependent simulation was performed until quasi-equilibrium textures were formed, after simulation times of $\ge 1 \mu s$. \ref{fig:struct1} shows the three different stationary smectic-A textures predicted by the model. Two general nano-scale structures are predicted, either a liquid crystalline defect core (\ref{fig:struct1}a and \ref{fig:struct1}b) or an isotropic/disordered core (\ref{fig:struct1}c) depending on the degree of undercooling. 

The fully smectic cores (\ref{fig:struct1}a and \ref{fig:struct1}b) are similar to the bipolar radial textures seen in confined nematics \cite{Yan2002}, where a pair of $+\frac{1}{2}$ disclinations form at the core to mediate the LC topology imposed by the combination of interfacial confinement geometry and anchoring. These orientational defects have biaxial core structures \cite{Schopohl1987}, which can be identified through the degree of biaxiality \cite{Kaiser1992,Kralj1999}:
\begin{equation} \label{eqn:biax}
\beta^2 = 1-6\frac{\left[\left(\bm{Q}\cdot\bm{Q}\right) : \bm{Q}\right]^2}{\left(\bm{Q} : \bm{Q}\right)^3}
\end{equation}
which ranges from $0 \rightarrow 1$ (no biaxiality to maximum biaxiality).  \ref{fig:struct1}a shows the smectic layer structure that is predicted by the model for intermediate and deep undercooling. The core is composed of an eccentric cylindrical/continuous smectic layer with the pair of $+\frac{1}{2}$ disclination (orientational) defects along the major axis. These texture and defect structures are  consistent with past work studying the transient growth of an initially radial-textured smectic-A spherulite \cite{Abukhdeir2008c}. Past results of cylindrical confinement using a lower order model \cite{Kralj1996} predicted the presence of a smectic $+1$ disclination, which was not observed to be stable under the conditions in this work.

\begin{figure*}[htp]
\centering
\subfigure[]{\includegraphics[width=0.3\linewidth]{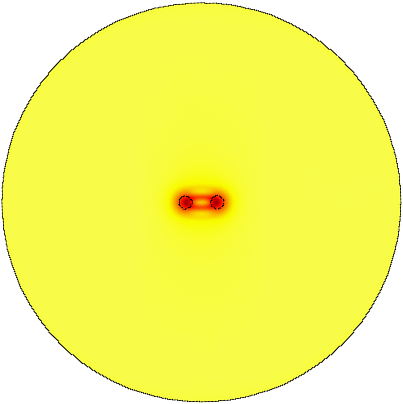}}
\subfigure[]{\includegraphics[width=0.3\linewidth]{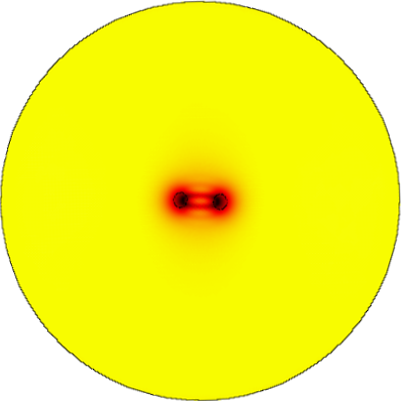}}
\subfigure[]{\includegraphics[width=0.3\linewidth]{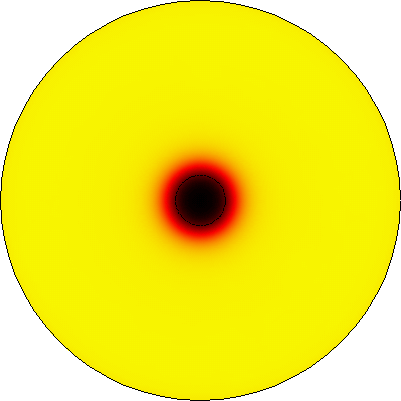}}\\
\subfigure[]{\includegraphics[width=0.3\linewidth]{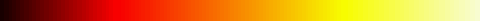}}
\caption{Simulation results of the smectic scalar order parameter $\psi$ (surface) and biaxiality (contour) $\beta^2$: (a) $T=331.00 K$, (b) $T=331.35 K$, and (c) $T=331.41 K$. Surface plot values range from $\psi=0 \rightarrow 0.161$ corresponding to (d) the color bar and the contour corresponds to $\beta^2=0.1$. \label{fig:struct2}}
\end{figure*}

\ref{fig:struct1}b shows the smectic layer structure as temperature is increased, where expansion of the smectic layer spacing collapses the cylindrical core smectic layer into a misoriented layer that undergoes a high degree of layer expansion at its center. \ref{fig:struct2}a and \ref{fig:struct2}b show the scalar smectic order parameter eqn. \ref{eq:smec_order_param} for both fully smectic core structures. These results agree with past work in nematic systems where an increase in temperature increases the separation distance between the disclination defects \cite{Yan2002}. In addition to defects in the underlying orientational order, the model predicts the formation of translational defects, elementary dislocations, enclosing the misoriented core layer. These observations elucidate structural differences, on the nano-scale, of cylindrical smectic domains.

The second general cylindrical domain structure (\ref{fig:struct1}a and \ref{fig:struct2}a) shows a fully melted/disordered core structure that has been predicted in the past for smectic-A filaments \cite{Rey2008a}. In this shallow undercooling scenario, the free energy difference between the smectic and isotropic phases is less than that of the formation of defects, as in the cases of \ref{fig:struct1}a-b.

\begin{figure*}[htp]
\centering
\subfigure[]{\includegraphics[height=2in]{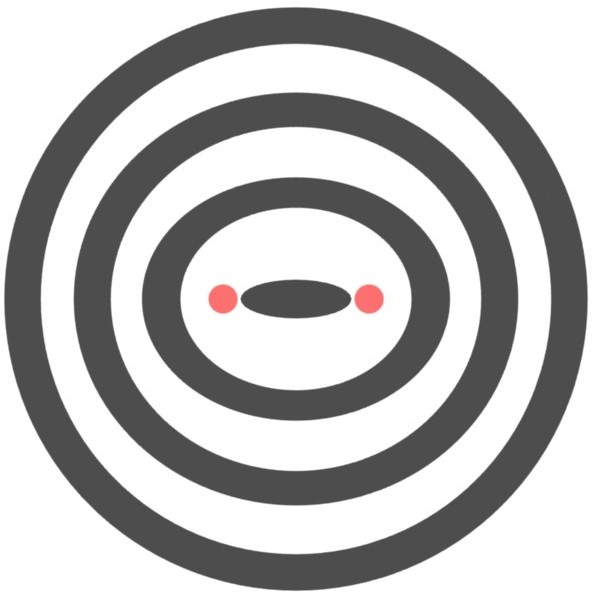}}
\subfigure[]{\includegraphics[height=2in]{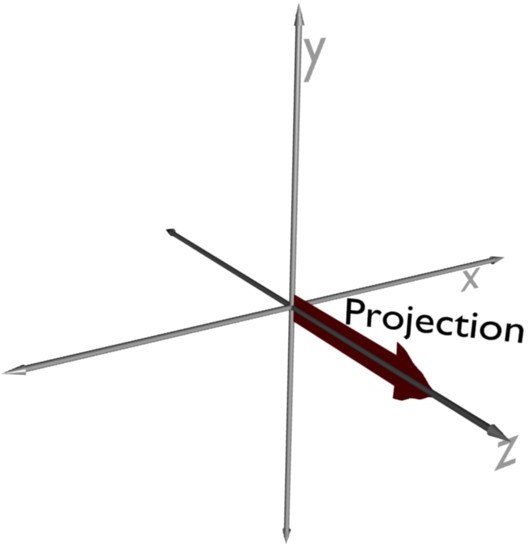}}
\subfigure[]{\includegraphics[height=2in]{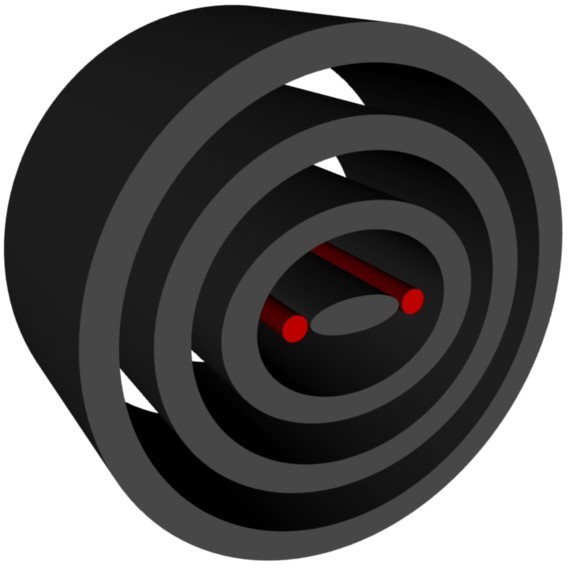}}\\
\subfigure[]{\includegraphics[height=2in]{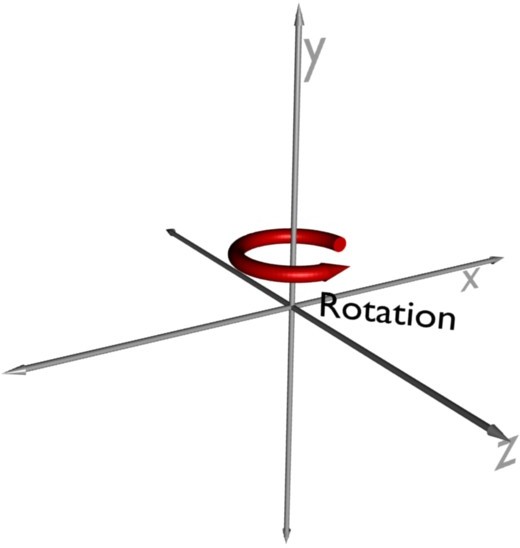}}
\subfigure[]{\includegraphics[height=2in]{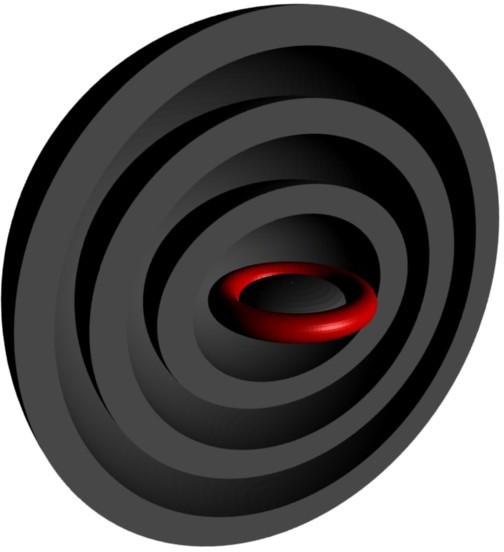}}
\subfigure[]{\includegraphics[height=2in]{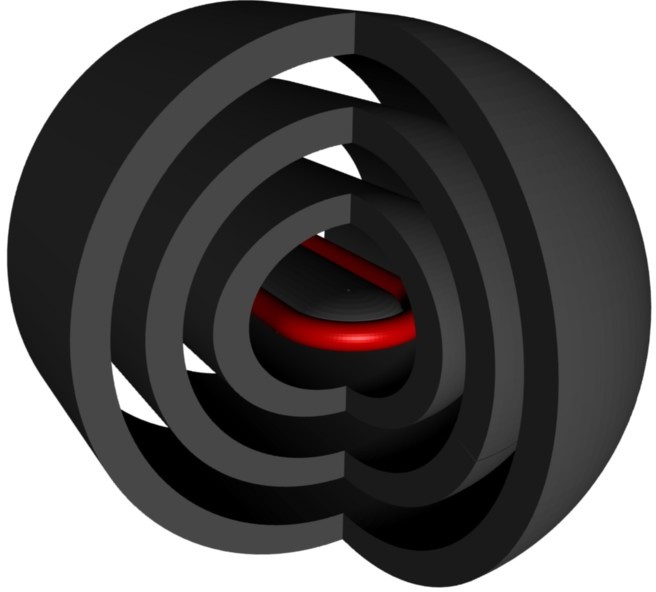}}
\caption{Schematics of (a) two-dimensional simulation results for intermediate/deep quenches \ref{fig:struct1}a (x/y/z-axes correspond to horizontal/vertical/out-of-plane axes) (b)-(c) the z-axis projection of the two-dimensional structure to a full three-dimensional cylindrical/filament domain (d)-(e) the y-axis rotation of the two-dimensional structure to a full three-dimensional spherical domain (f) sphero-cylindrical combination of z-axis projection (c) and y-axis rotation (e). \label{fig:symm}}
\end{figure*}

The two-dimensional simulation results presented also provide approximations of fully three-dimensional spheroidal and sphero-cylindrical domain structure. These three-dimensional geometries share symmetries with the presently simulated two-dimensional geometry and thus through projections on and rotations around these symmetry axes full three-dimensional nano-scale structure can be inferred. \ref{fig:symm} shows schematics of rotational axes of symmetry between a two-dimensional circle (present simulation geometry) and the extrusion/rotation relations with cylinders, spheres, and sphero-cylinders. 

The first relation of the two-dimensional domain \ref{fig:symm}a and \ref{fig:symm}b to a three-dimensional filament \ref{fig:symm}b involves a projection into the z-axis. This relation takes advantage of the two-dimensional simulation assumption of gradients in the z-axis equaling zero. Using the nano-scale structure predicted in \ref{fig:struct1}a, a direct quantitative filament structure is predicted to be composed of a eccentric cylindrical smectic-A layer capped by $+1/2$ disclination lines. The structure predicted at intermediate temperatures \ref{fig:struct1}b results in a disoriented smectic layer capped by both disclinations and elementary dislocation lines.

The second relation of the two-dimensional domain \ref{fig:symm}a to a three-dimensional spheroid \ref{fig:symm}b involves a rotation on the x/y-axis. This relation provides only qualitative information on the possible three-dimensional spherical domain structure. A direct qualitative relationship would require two-dimensional axial-symmetric simulation where the model is reformulated in spherical coordinates and subsequent gradients in the azimuthal axis are assumed zero. Past work simulating nematic materials using both two-dimensional rectangular (the present approach) and axial-symmetric formulations has shown qualitative structural agreement \cite{Fukuda2001,Fukuda2002}, supporting the present assumption. Thus, these rotations result in two possible smectic layer structures with either point or line defects (high-energy point defect structures are neglected \cite{Fukuda2002}). Using the nano-scale structure predicted in \ref{fig:struct1}a, the resulting spheroid structure is composed of a prolate spherical smectic-A layer with a single circular $+1/2$ disclination loop about its major axis. The structure predicted at intermediate temperatures \ref{fig:struct1}b results in a disoriented disc-shaped smectic-A layer confined by single circular disclination and elementary dislocation loops.

Finally, the sphero-cylindrical structure \ref{fig:symm}f which results from a combination of the structures inferred from relations shown in \ref{fig:symm}b and \ref{fig:symm}d involves capping smectic-A filaments with half-spherical domains. In these cases, defect loops are elliptical instead of circular.

\section{Conclusions}

In conclusion, a combined experimental/computational study of phase equilibrium and growth morphologies of novel PDLC materials which exhibit a direct isotropic/smectic-A transition was performed. Partial phase diagrams were determined for two different smectic PDLC materials: PS/10CB and PS/12CB. The approach of Benmouna et al \cite{Benmouna2000} utilizing the Flory-Huggins theory of isotropic mixing and Maier-Saupe-McMillan theory for smectic-A liquid crystalline ordering was used to computationally determine phase diagrams for both systems. The resulting computationally determined phase diagrams were found to be in good agreement with the experimental results, and provide both an estimation of the full phase diagram and the ability to predict the effects of using mono-disperse PS of different molar masses \cite{Benmouna2000}.

In addition to thermodynamic data, specific trends of the phase transition morphologies and dynamics were found for different compositions/phase-transition sequences and a general theory introduced identifying three distinct morphological regimes. These regimes include the formation of essentially spherical domains (isotropic/isotropic phase separation, ``stage 1''), highly anisotropic domains (isotropic/smectic-A phase separation, ``stage 2''), and sub-micron size spheroid domains (late isotropic/smectic-A phase separation, ``stage 3''). 

Finally, two-dimensional simulation of a high-order Landau-de Gennes model was used to determine nano-scale structure of generalized stage 1-2 morpologies. A parametric study in temperature was performed under three different degrees of undercooling. These novel results for two-dimensional smectic spherulites are consistent with past work in nematics \cite{Yan2002}, where the core structure is composed of $+1/2$ disclination loops. In addition to agreeing with past results, the presented simulations predict the smectic-A layer structure of the core to both include (intermediate undercooling) and lack (deep undercooling) elementary dislocation loops. Symmetry relations were then used to relate two-dimensional simulation results to fully three-dimensional experimentally observed morphologies, predicting possible nano-scale structures of spheroidal and sphero-cylindrical domains.

This study provides a basis for future experimental and theoretical study of these novel smectic PDLC materials which exhibit a direct isotropic/smectic-A phase transition. The next step of experimental analysis (using the techniques of TOM/POM) involve increased resolution studies of quench dynamics and growth morphologies, guided by the presented phase diagrams (\ref{fig:exp_phase_diag}a,b). In addition to experimental approaches, there is great promise in modeling and simulation of these systems, where length and time scales (nano-scale) inaccessible experimentally are able to be resolved. In particular, the extension of presently used Landau-de Gennes type model of Mukherjee, Pleiner, and Brand \cite{Mukherjee2001} to take into account the presence of polymers would be a substantial first step.

\section*{Acknowledgment}

This work was supported by a grant from the Natural Science and Engineering Research Council of Canada (NSERC) and a grant from the National Research Council of Argentina. Additional thanks to Dr. Diana Fasce (Institute of Materials Science and Technology, University of Mar del Plata) for training NMA on the use of DSC equipment.

\bibliographystyle{unsrt}
\bibliography{/home/nasser/Documents/references/references}

\end{document}